\documentclass[journal,final,twoside]{IEEEtran}
\usepackage{cite}

\ifCLASSINFOpdf
   \usepackage[pdftex]{graphicx}
   \graphicspath{{./figures/}{./pdf/}{./jpeg/}}
   \DeclareGraphicsExtensions{.pdf,.jpeg,.png}
\else
   \usepackage[dvips]{graphicx}
   \graphicspath{{./eps/}}
   \DeclareGraphicsExtensions{.eps}
\fi
\usepackage{algorithm}
\usepackage{algpseudocode}

\usepackage[cmex10]{amsmath}
\usepackage{amssymb}

\usepackage[caption=false,font=footnotesize]{subfig}
\usepackage{stfloats}

\usepackage{multirow}

\usepackage{footnote}

\usepackage{threeparttable}

\usepackage[cmex10]{amsmath}
\usepackage{amssymb}
\usepackage{bm}
\newcommand{\errcost}{{MSE}} 

\newcommand{\offseti}{{\gamma _i}}
\newcommand{\centeri}{{t_i}}

\renewcommand{\vec}[1]{\mathbf{#1}}

\newcommand{\vecslopei}{{\bm{\beta} _i}}
\newcommand{\vecslopeiTrans}{{\bm{\beta} _i^T}}
\newcommand{\veccenteri}{{\vec{x}_i}}
\newcommand{\signaljacobian}[2]{\nabla\varphi_{\left({#2}\right)}}
\newcommand{\signalPartialDerivative}[3]{{\frac{\partial\varphi\left({\vec{#1}}\right)}{\partial{{#1}_{#3}}}\Bigr|_{{\vec{#1}}={#2}}}}

\newcommand{\norm}[1]{\left\lVert#1\right\rVert}

\usepackage{varwidth}

\begin{document}
%
\title{On High-Resolution Adaptive Sampling\\of Deterministic Signals}
%
%
%

\author{Yehuda Dar and Alfred M. Bruckstein
\\
\thanks{The authors are with the Department of Computer Science, Technion, Israel. E-mail addresses: \{ydar,~freddy\}@cs.technion.ac.il.}
}

%
%

\markboth{}%
{~}
%



\maketitle

\begin{abstract}

In this work we study the topic of high-resolution adaptive sampling of a given deterministic signal and establish a connection with classic approaches to high-rate quantization.
Specifically, we formulate solutions for the task of optimal high-resolution sampling, counterparts of well-known results for high-rate quantization. 
Our results reveal that the optimal high-resolution sampling structure is determined by the density of the signal-gradient energy, just as the probability-density-function defines the optimal high-rate quantization form.
This paper has three main contributions: the first is establishing a fundamental paradigm bridging the topics of sampling and quantization.
The second is a theoretical analysis of nonuniform sampling relevant to the emerging field of high-resolution signal processing.
The third is a new practical approach to nonuniform sampling of one-dimensional signals that enables reconstruction based only on the sampling time-points and the signal extrema locations and values. Experiments for signal sampling and coding showed that our method outperforms an optimized tree-structured sampling technique.
\end{abstract}


\begin{IEEEkeywords}
High-resolution sampling, adaptive sampling, high-rate quantization.
\end{IEEEkeywords}

%
\IEEEpeerreviewmaketitle

\section{Introduction}
\IEEEPARstart{S}{ampling} and quantization are fundamental processes in signal digitization and coding techniques. 
Each of them is a field of research rich in theoretical and practical studies.
Quantization addresses the problem of discretizing a range of values by a mapping function, usually based on decomposition of the range into a finite set of non-intersecting regions.
Sampling a given signal, defined over a continuous and bounded domain, is the task of discretizing the signal representation (for example, representing a finite-length one-dimensional signal as a vector). In this paper we consider nonuniform sampling that relies on segmentation of the signal domain into non-overlapping regions, each represented by a single scalar coefficient (we do not consider here generalized sampling that uses projections of the signal onto a discrete set of orthonormal functions). 

\begin{figure*}[]
	\centering
	{\subfloat[]{\label{fig:general_sampling_reconstruction_procedure__theory}\includegraphics[height=0.15\textwidth]{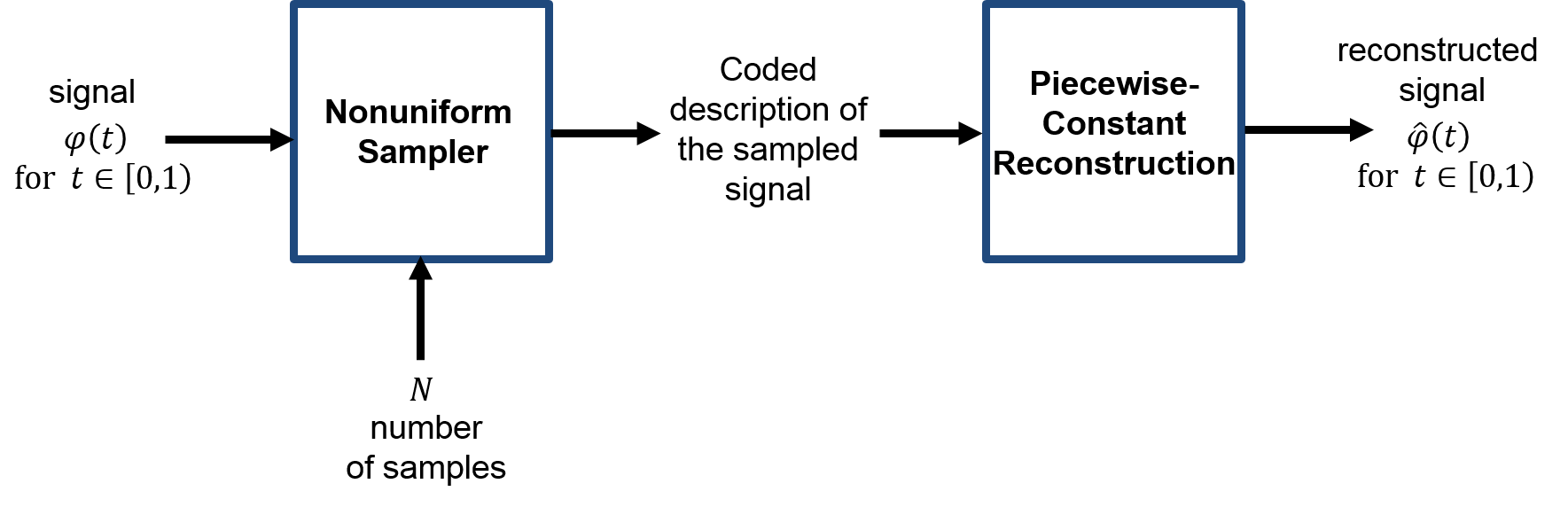}}}
	\\
	{\subfloat[]{\label{fig:general_sampling_reconstruction_procedure__practice}\includegraphics[height=0.15\textwidth]{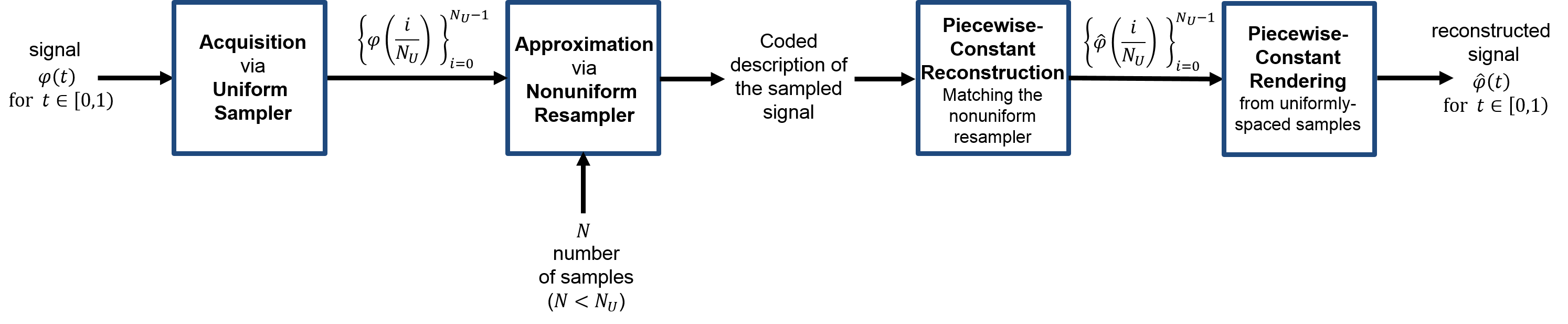}}}
	\caption{Nonuniform sampling of a deterministic signal. (a) The theoretic framework. (b) The corresponding practical settings. } 
	\label{Fig:general_sampling_reconstruction_procedure}
\end{figure*}

The signals considered in this work are deterministic in the sense that they are fully accessible to the sampler for its operation -- this concept appears differently in the theoretic and practical frameworks, as explained next. Our main analytic settings address a given one-dimensional signal defined over the continuous interval $ \left[ 0,1 \right) $, and its representation using a piecewise-constant approximation that relies on a high-resolution nonuniform segmentation (see Fig. \ref{fig:general_sampling_reconstruction_procedure__theory}). Obviously, in an acquisition process where, by definition, the continuous signal to-be-sampled is a-priori unknown, one cannot assume that the signal is readily available over $ \left[ 0, 1 \right) $. Nevertheless, the method we develop based on the ideal settings is effective for the practical architecture described in Fig. \ref{fig:general_sampling_reconstruction_procedure__practice}: an unknown continuous-time signal is first acquired using a very high-resolution uniform sampling, producing a discrete signal of $ N_U $ samples that goes through a resampler re-encoding the discrete signal using nonuniform segmentation with $ N < N_U $ breakpoints. The description produced by the nonuniform resampler then enables a piecewise-constant approximation of the sampled discrete signal and also a piecewise-constant continuous-time approximation of the original signal over $ \left[ 0,1 \right) $.

Historically, the discretizations in quantization and sampling were first implemented in their simplest forms relying on uniform divisions of the respective domains.
Then, the quantizer designs progressed to utilize nonuniform structures, exploiting input-data statistics to improve rate-distortion performance.
In his fundamental work, Bennett \cite{bennett1948spectra} suggested to implement nonuniform scalar quantization based on a companding model -- where the input value goes through a nonlinear mapping (compressor), the obtained value being uniformly quantized and then mapped back via the inverse of the nonlinearity (expander).
Moreover, under high-rate assumptions, Bennett derived a formula for approximating the quantizer distortion based on the source probability-density-function and the derivative of the nonlinear compressor function. This important formula is often referred to as Bennett's integral.
Bennett's work was followed by a long line of theoretic and algorithmic studies of the nonuniform quantization problem.
A prominent branch of research addressed the scenario of nonuniform quantization at high-rates (for example, see \cite{panter1951quantization,gersho1979asymptotically}), where the quantizer has a large number of representation-values to be wisely located at the quantizer-design stage.
The popularity of high-rate studies is due not only to their relevance in addressing high-quality coding applications, but also to the possibility to gain useful theoretical perspectives. Specifically, reasonable assumptions made for high-rate quantization often led to convenient closed-form mathematical solutions that, in turn, provided deep insights into rate-distortion trade-offs.


Sampling has been prevalently studied for the purpose of acquiring signals based on global or coarse characterizations such as their bandwidth. The classical uniform sampling theorems (see a detailed review in \cite{jerri1977shannon}) were also extended to the nonuniform settings where the variable sampling-rate is adapted to, e.g., the signal's local-bandwidth estimate \cite{horiuchi1968sampling,clark1985transformation,brueller1998non,wei2007sampling}. In \cite{clark1985transformation,brueller1998non,wei2007sampling}, a nonuniform sampling grid was designed by mapping the signal's time axis to expand portions corresponding to high local-bandwidth at the expense of segments containing content of lower local-bandwidth. After applying this transformation, signal acquisition could subsequently be carried out using  uniform sampling.
While these works refer to acquiring signals in real time, we consider sampling of fully-accessible deterministic signals for representation-oriented tasks such as compression. 
Despite the above discussed essential difference in the sampling and reconstruction procedures, some results presented in this paper conceptually resemble sampling methods for acquisition of unknown signals using coordinate transformation. The results presented here for sampling given deterministic signals may also be interpreted as complements to the adaptive sampling paradigms previously proposed for signal acquisition. In our deterministic settings, the analysis and sampling-rate design directly rely on properties obtained from the given signal instead of coarse, local average-based, spectral characterizations.

Practical signal compression (see, e.g., \cite{sullivan2012overview,shukla2005rate}) usually considers a digital high-resolution input function that is going through a nonuniform adaptive partitioning of the domain followed by a quantization procedure that suits the partition size and shape. 
Analytically, the digital input can be regarded as a signal defined over a continuous domain that needs to be nonuniformly sampled and quantized.
In practical coding procedures, the nonuniform domain segmentation often relies on structured partitions of the domain in order to reduce the bit-cost and computational complexity (examples for the common use of tree-structured partitioning, e.g., defined based on binary or quad trees, are available in \cite{shusterman1994image,shukla2005rate}).
This application is one of the main motivations to the study of adaptive sampling of deterministic signals. Nevertheless, we consider here the nonuniform sampling problem in its most general form in order to provide insights to the basic sampling problem, that may be useful to problems beyond compression.
For example, adaptive sampling is often used in computer graphics in the task of Halftoning (e.g., \cite{ulichney1999review,lau2006blue}) and in rendering an image from a 2D/3D-graphical model, where the nonuniform sampling pattern is described by point distributions (e.g., \cite{mitchell1987generating,mitchell1991spectrally,mccool1992hierarchical,fattal2011blue,zhong2016kernel}).
It should be noted that the signal processing and computer graphics contexts of the sampling problem are quite different. 
Interestingly, it was suggested in \cite{mccool1992hierarchical} to practically employ the Lloyd algorithm \cite{lloyd1982least}, originally intended for quantizer design, for improving the sampling point distribution.
Recent studies \cite{fattal2011blue,zhong2016kernel} extended the point-distribution computation to rely on kernel functions, so that the resulting task resembles a generalized signal-sampling procedure.
While the conceptual relation between vector-quantization and data-representation has been understood and used in graphics applications, we argue that a clear mathematical analysis that demonstrates the connection between quantization and high-resolution signal-sampling has not been provided yet.

Some recent papers \cite{neuhoff2013information,kipnis2014gaussian,kipnis2016distortion} explore the sampling task from a stochastic, information theoretic rate-distortion tradeoff perspective, considering lossy compression of the sampled values. Their focus is on uniform \cite{neuhoff2013information,kipnis2016distortion} and nonuniform \cite{kipnis2014gaussian} sampling where the sampling design is based on spectral characteristics of stationary signals. Another interesting direction of nonuniform sampling was explored in \cite{feizi2010locally,feizi2012time} where the sampling intervals are recursively determined based on previously recorded data. This process does not require the coding of the sampling intervals. The sampling design in \cite{feizi2010locally,feizi2012time} is based on either the local properties of the stochastic process whose realizations are sampled, or the local behavior of deterministic signals exhibited in their Taylor expansion.
Another attractive nonuniform sampling approach was introduced lately in \cite{martinez2018delta}, defining the samples based on the crossings of uniformly-spaced signal-amplitude levels thereby enabling a signal representation using only the sampling time points. This remarkable general idea of describing nonuniformly sampled data by the time-points also appears in our practical sampling method, however, our high-resolution case leads to significant conceptual differences with respect to \cite{martinez2018delta}, as will be explained later in this paper.

In this paper we theoretically and practically explore the task of high-resolution adaptive sampling of a given deterministic signal.
The sampling analysis provided emerges from ideas similar to the ones that were applied to the study of high-rate quantization \cite{gray1998quantization}, thereby linking sampling and quantization in a new and enlightening way.
We analytically formulate the optimal high-resolution sampling of one-dimensional signals, based on the Mean-Squared-Error (MSE) criterion, showing that the optimal partitioning is determined by the cube-root of the signal-derivative energy. This result corresponds to the work by Panter and Dite \cite{panter1951quantization}, where the optimal one-dimensional quantizer is designed based on the cube-root of the probability-density-function. We also connect this result to the fundamental analysis given by Bennett \cite{bennett1948spectra} for high-rate nonuniform quantization based on companding.

We rely on our analytical findings to establish a practical sampling method for one-dimensional signals, and show its effectiveness for coding of analytic and audio signals. 
The fact that the proposed sampling is based on an arbitrary segmentation of the time axis may be suspected to lead to an inevitably high representation cost. However, we utilize the underlying properties of our optimized nonuniform partitioning where, by our companding-based construction, all the segments have an equal accumulated amount of the cube-root of the signal-derivative energy. The latter means that given a segment length, assuming the signal is monotonic within the segment, we can infer the local average-derivative of the signal. We present a procedure for sampling non-monotonic signals coupled with a sequential reconstruction process requiring only the sampling times, the signal extrema times and amplitudes, and few additional global details having a marginal cost. Experiments showed that our strategy outperforms an optimized adaptive tree-based sampling technique.

We continue our theoretical study by addressing the problem of sampling $ K $-dimensional signals. We show that the optimal sampling-point density is determined by the density of the $ \frac{K}{K+2} $-power of the signal-gradient energy, a generalization of the one-dimensional result. We obtain this result based on assumptions that parallel a famous conjecture given by Gersho \cite{gersho1979asymptotically} in his analysis of high-rate quantization.
Gersho's conjecture states that, for asymptotically high rate, the optimal $ K $-dimensional quantizer is formed by regions that are approximately congruent and scaled versions of a $ K $-dimensional convex polytope that optimally tessellates the $ K $-dimensional space (where the polytope optimality is in the sense of minimum normalized moment of inertia \cite{gersho1979asymptotically}).
The latter holds for a given $ K $ only if the optimal tessellation of the $ K $-dimensional space is a lattice, and therefore constructed based on a single optimal polytope.
This assumption significantly simplifies the explicit calculation of the quantizer's distortion.
This conjecture draws its credibility from two prominent sources. 
First, it is known that the best tessellation is a lattice for $ K=1 $ (based on equal-sized intervals) and for $ K=2 $ (based on the hexagon shape \cite{newman1982hexagon}). 
While not proven yet for $ K=3 $, it is also believed that the optimal three-dimensional tessellation is the body-centered cubic lattice \cite{barnes1983optimal,du2005optimal}. 
Second, Gersho's distortion formula conforms to the structure of the expression rigorously obtained by Zador \cite{zador1982asymptotic}. Moreover, Gersho's conjecture determines the value of the multiplicative-constant (left unspecified) in Zador's formula. 
Hence, the possible inaccuracy in Gersho's conjecture will affect only the multiplicative constant, and the deviation is assumed to be moderate \cite{gray1998quantization}.
Due to this, the conjecture, still unproved for $ K\ge 3 $, is widely considered as a valuable tool for analysis of high-rate quantization (see the thorough discussion in \cite{gray1998quantization}).

The analysis provided in this paper to high-resolution multidimensional sampling is based on two main assumptions. First, the signal is assumed to be approximately linear within each of the sampling regions. Second, sampling regions are assumed to be approximately congruent and scaled forms of the optimal $ K $-dimensional tessellating convex polytope -- just as in Gersho's conjecture for high-rate quantization. 
These high-resolution assumptions yield our main result that the optimal sampling-point density is determined by the density of the $ \frac{K}{K+2} $-power of the signal-gradient energy.
We emphasize the importance of the signal's local-linearity assumption as a prerequisite stage that mathematically connects the signal-sampling problem to the conjecture on the high-resolution cell arrangement.

This paper is organized as follows. In section \ref{sec:Analysis for One-Dimensional Signals} we mathematically analyze the optimal sampling of one-dimensional signals, and demonstrate it numerically. 
In section \ref{sec:Practical Sampling Method} we present a practical sampling method, relying on our theoretic results, and experimentally exhibit its utilization for signal compression. 
In section \ref{sec:Analysis for Multidimensional Signals} we generalize our study by theoretically addressing the optimal sampling of multidimensional signals.
Section \ref{sec:Conclusion} concludes this paper.

\section{Analysis for One-Dimensional Signals}
\label{sec:Analysis for One-Dimensional Signals}

\subsection{Optimal High-Resolution Sampling}
\label{subsec:Optimal High-Resolution Sampling}

Let us consider a one-dimensional signal $ \varphi \left(t\right) $ defined as a differentiable function 
\begin{IEEEeqnarray}{rCl}
	\label{eq:continuous signal}
	\varphi : \left[0,1 \right) \rightarrow \left[\varphi _L, \varphi _H \right],
\end{IEEEeqnarray}
defined for $ t $ in the interval $ \left[0,1 \right) $ and having values from a bounded range $ \left[\varphi _L, \varphi _H \right] $.
This signal is sampled based on its partition to $ N \in \mathbb{N} $ non-overlapping variable-length segments, where the $ i^{th} $ subinterval $ \left[a_{i-1},a_{i}\right) $ is associated with the sample $ \varphi_i $ for $ i=1,...,N $. 
We assume a segmentation structure satisfying $ a_{0}=0 $, $ a_{N}=1 $, and $ a_{i-1}<a_{i} $ for $ i=1,...,N $.
The sampling procedure is coupled with a reconstruction that provides the continuous-time piecewise-constant signal
\begin{IEEEeqnarray}{rCl}
	\label{eq:reconstructed continuous signal}
	\hat\varphi \left( t \right) = {\varphi _i} \text{~~~~for~~~~}  t\in \left[a_{i-1},a_i\right).
\end{IEEEeqnarray}
The sampling is optimized to minimize the mean-squared-error (MSE), expressed as
\begin{IEEEeqnarray}{rCl}
	\label{eq:sampling MSE}
	\errcost \left( {\left\{ {{a_i}} \right\}_{i = 1}^{N - 1},\left\{ {{\varphi _i}} \right\}_{i = 1}^{N}} \right) = \mathop \sum \limits_{i = 1}^N \mathop \int \limits_{{a_{i - 1}}}^{{a_i}} {\left( {\varphi \left( t \right) - \varphi _i} \right)^2}dt \nonumber\\ 
\end{IEEEeqnarray}
exhibiting the roles of the signal partitioning and sample values. Note that in (\ref{eq:sampling MSE}), as also in this entire section, averaging over the unit-interval length is implicit.

Optimizing the sampling coefficients, $ \left\lbrace {\varphi _i} \right\rbrace _{i=1}^N $, given a partitioning $ \left\lbrace {a _i} \right\rbrace _{i=0}^{N} $ is a convex problem that can be analytically solved to show that the optimal $ i^{th}  $ sample is the signal average over the corresponding subinterval, namely,
\begin{IEEEeqnarray}{rCl}
	\label{eq:optimal sample}
	{\varphi _i ^{opt}} = \frac{1}{\Delta _i} \mathop \int \limits_{{a_{i - 1}}}^{{a_i}} \varphi \left( t \right)dt 
\end{IEEEeqnarray}
where $ \Delta_i \triangleq a_i - a_{i-1} $ is the length of the $ i^{th} $ subinterval.

We continue the analysis by assuming high-rate sampling, meaning that $ N $ is large enough to result in small sampling-intervals that, however, may still have different lengths.
Furthermore, the sampling intervals are assumed to be sufficiently small such that, within each of them, the signal is well approximated via a local linear form -- an argument that is analyzed next.
Accordingly, for $ t \in \left[a_{i-1},a_i\right) $ ($ i = 1,...,N $), we consider the first-order Taylor approximation of the signal about the center of the $ i^{th} $ sampling interval, $ \centeri \triangleq \frac{1}{2}\left(a_{i-1}+a_i\right) $,
\begin{IEEEeqnarray}{rCl}
	\label{eq:first-order Taylor approximation}
	\varphi (t) = \varphi \left(\centeri\right) + \varphi ' \left(\centeri\right) \cdot \left(t-\centeri\right) + o\left(\lvert t-\centeri\rvert\right)
\end{IEEEeqnarray}
where the remainder term $ o\left(\lvert t-\centeri\rvert\right) $ corresponds to our high-resolution assumption in describing the approximation error for $ t \rightarrow t_i $. 

Using the linear approximation, and by (\ref{eq:optimal sample}), the optimal sample in the $ i^{th} $ subinterval is given by 
\begin{IEEEeqnarray}{rCl}
	\label{eq:optimal sample - approximation following linearization}
	{\varphi _i ^{opt}} = \varphi \left(\centeri\right) + o\left(\Delta_i \right)
\end{IEEEeqnarray}
due to the fact that the average of a linear function over an interval is its value at the interval's center. The term $ o\left(\Delta_i \right) $ describes the error in the sample value due to the linear approximation. The size of the error term as $ \Delta_i \rightarrow 0 $ is provided in Appendix \ref{appendix:Analysis of Inaccuracies Due to the Signal Linearity Assumption} and relies on assuming that the second derivative of the signal exists and bounded over the sampling interval.
Then, the MSE of the $ i^{th} $ subinterval is expressed as 
\begin{IEEEeqnarray}{rCl}
	\label{eq:sampling MSE of subinterval }
	\errcost _i \left( a_{i-1}, a_i \right) & = & \frac{1}{\Delta _i} \mathop \int \limits_{{a_{i - 1}}}^{{a_i}} {\left( {\varphi \left( t \right) - \varphi _i^{opt}} \right)^2}dt
	\\ \label{eq:sampling MSE of subinterval - function of derivative}
	& = & \frac{1}{12} \left({\varphi ' \left(\centeri\right)}\right) ^2 {\Delta _i^2} + o\left( \Delta_i^2 \right)
\end{IEEEeqnarray}
revealing the effect of signal-derivative energy on the sampling MSE. The derivation of (\ref{eq:sampling MSE of subinterval - function of derivative}) is detailed in Appendix \ref{appendix:Analysis of Inaccuracies Due to the Signal Linearity Assumption}.

Returning to the total sampling MSE, corresponding to optimal coefficients, and relying on its relation to the subintervals MSE yields (recall the implicit normalization to unit-interval length)
\begin{IEEEeqnarray}{rCl}
	\label{eq:total sampling MSE - average of subinterval MSE values}
	\errcost \left( {\left\{ {{a_i}} \right\}_{i = 1}^{N - 1} }\right) & = & \mathop \sum \limits_{i = 1}^N  {\Delta _i \errcost _i \left( a_{i-1}, a_i \right) } \nonumber
	\\ \label{eq:total sampling MSE - function of derivative}
	& = & \frac{1}{12} \mathop \sum \limits_{i = 1}^N  { \left({\varphi ' \left(\centeri\right)}\right) ^2  {\Delta _i^3} } + o\left( \Delta_{\text{max}}^3 \right) \nonumber\\
\end{IEEEeqnarray}
where $\Delta_{\text{max}} \triangleq \max\lbrace \Delta_1 ,... ,\Delta_N\rbrace$ is the largest subinterval. Accordingly, the inaccuracies in evaluating the MSE using the signal linearity assumption are of size $ o\left( \Delta_{\text{max}}^3 \right) $.

Let us connect our discussion to the classical approach of studying high-rate quantization based on the reproduction-value density function (see examples in \cite{lloyd1982least,gersho1979asymptotically,na1995bennett,gray1998quantization}).
Following our scenario of high-resolution sampling we assume that the sampling-point layout can be described via a sampling-point density function, $ \lambda  \left( t \right) $, such that a small interval of length $ \bar\Delta $ around $ \bar t $ approximately contains $ \bar\Delta \cdot \lambda \left( \bar t \right)$ sampling points. Moreover, the sampling-point density is related to the sampling intervals via 
\begin{IEEEeqnarray}{rCl}
		\label{eq:sampling point density - definition - one-dimensional signals}
	\lambda \left( t \right) \approx \frac{1}{N\cdot \Delta_i}   ,  \text{~~~~for~~}  t\in \left[ a_{i-1}, a_i\right) .
\end{IEEEeqnarray}
As we consider arbitrarily large values of $ N $, the density $ \lambda \left( t \right) $ is assumed to be a smooth function.

Then, plugging the relation $ \Delta_i \approx 1/\left(N \cdot\lambda \left( t_i \right)\right) $ into the sampling-error expression (\ref{eq:total sampling MSE - function of derivative}), in addition to approximating the sum as an integral and omitting the explicit inaccuracy term, yields
\begin{IEEEeqnarray}{rCl}
	\label{eq:Bennet integral for sampling}
	\errcost \left(  \lambda  \right) \approx \frac{1}{12 N^2} \mathop \int \limits_{0}^{1}  \frac{ \left({\varphi ' \left(t \right)}\right) ^2}{\lambda ^2 (t)} dt.
\end{IEEEeqnarray}
Here the sampling structure and the resulting MSE are determined by the sampling-point density $ \lambda(t) $.
The last MSE expression can be interpreted as the sampling equivalent of Bennett's integral for nonuniform quantization \cite{bennett1948spectra}. Commonly, the expression form in (\ref{eq:Bennet integral for sampling}) is minimized via H\"{o}lder's inequality (see examples in \cite{gersho1979asymptotically,na1995bennett,gray1998quantization}). For our problem, see details in Appendix \ref{appendix:Sampling-Point Density Optimization - one dimensional}, we have the optimal sampling-point density given by 
\begin{IEEEeqnarray}{rCl}
	\label{eq:sampling point density - one-dimensional signals - optimal - continuous}
	{\lambda^{opt}} \left( t \right) = \frac{\sqrt[3]{ \left({\varphi ' \left(t\right)}\right) ^2  }}{\mathop \int \limits_{0}^1 { \sqrt[3]{ \left({\varphi ' \left(z\right)}\right) ^2  }dz}} , 
\end{IEEEeqnarray}
and the optimal sampling MSE is 
\begin{IEEEeqnarray}{rCl}
	\label{eq:optimal sampling MSE}
	\errcost \left( {\lambda^{opt}} \right) \approx \frac{1}{12 N^2} \left( \mathop \int \limits_{0}^1 { \sqrt[3]{ \left({\varphi ' \left(t\right)}\right) ^2  }dt} \right)^{3} . 
\end{IEEEeqnarray}
The error expression in (\ref{eq:optimal sampling MSE}) is a product composed of two parts: sampling error for a simple linear signal (with a unit slope), and a term expressing the nonlinearity of the given signal based on its derivative energy. Moreover, H\"{o}lder's inequality also shows that (\ref{eq:optimal sampling MSE}) is the global minimum.
The MSE expression in (\ref{eq:optimal sampling MSE}) is the sampling counterpart of the famous Panter-Dite formula for high-rate quantization MSE \cite{panter1951quantization}. Evidently, while the quantization derivations were determined by the probability-density-function of the source, the sampling analysis provided here depends on the signal-derivative energy.

Using the sampling-point density, $ \lambda(t) $, we can implement our nonuniform sampling via the companding design. Companding \cite{bennett1948spectra} is a widely-known technique for implementing nonuniform quantization based on a uniform quantizer. This is achieved by applying a nonlinear compressor function on the input value, then applying uniform quantization and mapping the result back via an expander function (the inverse of the compressor).
Since in sampling we address the problem of discretizing the signal domain, the corresponding compressor and expander functions operate on the signal domain (i.e., as a nonlinear scaling of the time axis). The optimal compressor function is defined based on the optimal density (\ref{eq:sampling point density - one-dimensional signals - optimal - continuous}) as
\begin{IEEEeqnarray}{rCl}
	\label{eq:optimal sampling compressor function}
	u\left(t\right) = \int\limits_{0}^{t} {\lambda^{opt}(z)dz} = \frac{\int\limits_{0}^{t} {\sqrt[3]{ \left({\varphi ' \left(z \right)}\right) ^2  } dz} }{\int\limits_{0}^{1} {\sqrt[3]{ \left({\varphi ' \left(z \right)}\right) ^2  } dz}} .
\end{IEEEeqnarray}
The corresponding expander, $ v\left(\tau\right) $, is the inverse function of the compressor, hence, it can be defined via the relation 
\begin{IEEEeqnarray}{rCl}
	\label{eq:optimal sampling expander function}
	\int\limits_{0}^{v\left(\tau\right)} {\lambda^{opt}(z)dz}  = \tau .
\end{IEEEeqnarray}
The last equation has a unique solution for strictly monotonic signals where the signal-derivative energy is positive over the entire domain, leading to a strictly-monotonic increasing compressor function (\ref{eq:optimal sampling compressor function}) that is invertible and, thus, defines the expander function. Two suggestions for the treatment of non-monotonic signals will be described next.

\subsubsection{Using a Strictly-Positive Extension of the Signal's Derivative-Energy}
\label{subsubsec:Strictly-Positive Extension}
~\\
Consider the expander function in (\ref{eq:optimal sampling expander function}) to construct the boundaries of the nonuniform segmentation of $ \left[0,1\right) $ via
\begin{IEEEeqnarray}{rCl}
	\label{eq:optimal sampling interval - boundaries - continuous}
	a_i^{opt} = v\left(\frac{i}{N}\right)  ~~~~,i=1,...,N,~
\end{IEEEeqnarray}
i.e., evaluating the inverse of the compressor function at $ N $ equally-spaced points. 
However, for non-monotonic signals (that also may have constant-valued segments), it is likely to need expander values at points where the compressor function is not invertible -- an issue that can be solved as follows. Since the problem occurs where the signal derivative is zero, we define the following extension of the derivative energy:
\begin{IEEEeqnarray}{rCl}
	\label{eq:extended derivative energy density - one-dimensional signals}
	g ^2 _{\varepsilon} \left( t \right) = \begin{cases}
		\left({ \varphi ' \left( t \right) }\right)^2  & \text{for }\left({ \varphi ' \left( t \right) }\right)^2 > \varepsilon\\
		\varepsilon  & \text{otherwise}
	\end{cases}		
\end{IEEEeqnarray}
where $ \varepsilon > 0 $ is an arbitrarily small constant. The corresponding extension of the optimal sampling-point density (\ref{eq:sampling point density - one-dimensional signals - optimal - continuous}) is 
\begin{IEEEeqnarray}{rCl}
	\label{eq:sampling point density - one-dimensional signals - optimal - continuous - extended}
	\lambda^{opt}_{\varepsilon} \left( t \right) = \frac{\sqrt[3]{ g ^2 _{\varepsilon} \left( t \right)  } }{\int\limits_{0}^{1} {\sqrt[3]{ g ^2 _{\varepsilon} \left( z \right)  } dz}}
\end{IEEEeqnarray}
Accordingly, the density $ \lambda^{opt}_{\varepsilon}(t) $ enables treatment of non-monotonic signals, while closely approximating the density $ \lambda^{opt}(t) $ in (\ref{eq:sampling point density - one-dimensional signals - optimal - continuous}). Replacing  $ \lambda^{opt}(t) $ with $ \lambda^{opt}_{\varepsilon}(t) $ in (\ref{eq:optimal sampling compressor function})-(\ref{eq:optimal sampling expander function}) provides a practically useful compressor-expander pair, in the sense that the compressor function is invertible everywhere in the domain, assuring the computations in (\ref{eq:optimal sampling interval - boundaries - continuous}).

\subsubsection{Sequential Solution via Integration Thresholds}
\label{subsubsec:Sequential Solution via Integration Thresholds}
~\\
Eq. (\ref{eq:optimal sampling expander function}) defines the segment level $ a_i^{opt} $ ($ i=1,...,N $) as 
\begin{IEEEeqnarray}{rCl}
	\label{eq:optimal sampling interval - boundaries - continuous - explicit}
	{\int\limits_{0}^{a_i^{opt}} {\lambda^{opt} \left(z\right)  } dz} = \frac{i}{N}
\end{IEEEeqnarray}
implying that $ a_i^{opt} $ can be evaluated given the former partitioning level $a_{i-1}^{opt}$ via 
\begin{IEEEeqnarray}{rCl}
	\label{eq:optimal sampling interval - boundaries - continuous - explicit - sequential}
	{\int\limits_{a_{i-1}^{opt}}^{a_i^{opt}} {\sqrt[3]{ \left({\varphi ' \left(t \right)}\right) ^2   } dt}} = \frac{1}{N} {\int\limits_{0}^{1} {\sqrt[3]{ \left({\varphi ' \left(z \right)}\right) ^2 } dz}}
\end{IEEEeqnarray}
i.e., we just need to continuously integrate $ {\sqrt[3]{ \left({\varphi ' \left(t \right)}\right) ^2}} $ starting at $t={a_{i-1}^{opt}}$ until we reach the threshold level defined by the right side of the last equation. Then, the time of achieving the threshold level defines ${a_{i}^{opt}}$. 
Since a partitioning level is placed at the (first) time the threshold level is obtained, intervals of zero signal derivative-energy do not cause ambiguity in segmentation definitions.
Importantly, one should note that the (optimal) threshold, defined according to Eq. (\ref{eq:optimal sampling interval - boundaries - continuous - explicit - sequential}) as 
\begin{IEEEeqnarray}{rCl}
	\label{eq:optimal sampling interval - sequential - optimal threshold}
	T_{opt} = \frac{1}{N} {\int\limits_{0}^{1} {\sqrt[3]{ \left({\varphi ' \left(z \right)}\right) ^2 } dz}},
\end{IEEEeqnarray}
requires knowing the signal derivative over the entire $ \left[0,1\right) $ interval. 
Note however that the segmentation levels are here well-defined for any given nonzero signal.

\subsection{The Sampling-Quantization Duality at High Resolution}
\label{subsec:The Sampling-Quantization Duality at High Resolution}

The above developments for optimal high-resolution sampling show a clear correspondence to well-known results from high-rate quantization studies. 
The connection between sampling and quantization, in the high-resolution settings, emerges from local linear approximation of the signal within each of the sampling intervals, implying a constant signal-derivative within each interval. This mirrors the assumption of locally constant probability density in the high-rate quantization problem.
In turn, our construction yields an optimal sampling procedure determined by the signal's derivative-energy density. Specifically, the MSE forms in (\ref{eq:Bennet integral for sampling}) and (\ref{eq:optimal sampling MSE}) and the corresponding companding approach are as in the classical high-rate quantization results, however, here they correspond to the signal's derivative-energy density instead of the probability density function of a source to be quantized.

In the sequel, let us consider the scalar quantizer design for a random variable $ X $, defined by the probability density function $ p_X \left(x\right) $ that may be positive only for $ x \in [x_L , x_H) $. Then, recall Bennett's integral \cite{bennett1948spectra} for the MSE of high-rate quantization using $ N $ reproduction values:
\begin{IEEEeqnarray}{rCl}
	\label{eq:Bennet integral for quantization}
	\errcost_Q \left(  \lambda_Q  ; p_X \right) \approx \frac{1}{12 N^2} \mathop \int \limits_{x_L}^{x_H}  \frac{p_X \left( x \right) }{\lambda_Q ^2 (x)} dx 
\end{IEEEeqnarray}
where, as interpreted in \cite{lloyd1982least}, $ \lambda_Q \left(x\right)$ is the reproduction-point density that defines the quantizer structure to optimize, and $p_X\left(x\right)$ is the given input probability density function. 

In this subsection we distinguish between optimizations originating in sampling and quantization procedures by the following notations: the quantizer design procedure addresses Bennett's integral (\ref{eq:Bennet integral for quantization}) to minimize the quantization distortion, $ MSE_Q $, as a function of the reproduction-point density $ \lambda_Q  $. Similarly, the sampling procedure considers the sampling counterpart of Bennett's integral presented in (\ref{eq:Bennet integral for sampling}), that using this subsection's notations is formulated as 
\begin{IEEEeqnarray}{rCl}
	\label{eq:Bennet integral for sampling - notations of quantization-sampling duality}
	\errcost_S \left(  \lambda_S  ; \varphi \right) \approx \frac{1}{12 N^2} \mathop \int \limits_{0}^{1}  \frac{ \left( \varphi ' \left( t \right) \right) ^2 }{\lambda_S ^2 (t)} dt ,
\end{IEEEeqnarray}
where $ MSE_S $, the distortion of sampling the signal $ \varphi(t) $, is to be minimized by optimizing the sampling-point density $ \lambda_S $.

\subsubsection{Optimal Sampling via Optimal Quantizer Design}
~\\
For a given differentiable signal, $ \varphi \left(t\right) $ defined for $ t\in[0,1) $, we can define a probability density function, $ p_{\varphi}(x) $, defined for $x\in[0,1) $ via 
\begin{IEEEeqnarray}{rCl}
	\label{eq:equivalent probability density function}
	 p_{\varphi}(x) = \frac{\left({\varphi ' \left(x \right)}\right) ^2  }{\int\limits_{0}^{1} { \left({\varphi ' \left(z \right)}\right) ^2  dz}} , 
\end{IEEEeqnarray}
i.e., the probability density is the signal's local squared-derivative normalized by the total derivative energy. Evidently, $ p_{\varphi}(x) $ is non-negative valued and integrates to one, hence, it is a valid probability density function. 

Bennett's integral for high-rate quantization of a source distributed according to $p_{\varphi}$ is obtained by plugging the definition of $ p_{\varphi}(x) $ from (\ref{eq:equivalent probability density function}) into (\ref{eq:Bennet integral for quantization}), resulting in 
\begin{IEEEeqnarray}{rCl}
	\label{eq:Bennet integral for quantization - relation to sampling MSE - part 1}
	\errcost_Q \left(  \lambda_Q ; p_{\varphi} \right) \approx \frac{1}{12 N^2  {\mathcal{E}_{\varphi'}} } \mathop \int \limits_{0}^{1}  \frac{ \left({\varphi ' \left(x \right)}\right) ^2  }{\lambda_Q ^2 (x)} dx.
\end{IEEEeqnarray}
where $ \lambda_Q\left(\cdot \right) $ is the reproduction-point density defining the high-rate quantizer structure, and the total derivative-energy of the signal is denoted here as 
\begin{IEEEeqnarray}{rCl}
	\label{eq:total derivative energy - short notation}
	\mathcal{E}_{\varphi'} \triangleq \int\limits_{0}^{1} { \left({\varphi ' \left(z \right)}\right) ^2  dz} . 
\end{IEEEeqnarray}
Then, the high-resolution sampling MSE, $ MSE_S $, formulated in (\ref{eq:Bennet integral for sampling - notations of quantization-sampling duality}) is related to the high-rate quantization MSE in (\ref{eq:Bennet integral for quantization - relation to sampling MSE - part 1}) via 
\begin{IEEEeqnarray}{rCl}
	\label{eq:Bennet integral for quantization - relation to sampling MSE}
	\errcost_Q \left(  \lambda_Q ; p_{\varphi} \right) \approx \frac{1}{ {\mathcal{E}_{\varphi'}} } \errcost_S \left(  \lambda_Q ; \varphi \right) ,
\end{IEEEeqnarray}
i.e., the MSE of quantizing a source defined by $p_{\varphi}$ using a quantizer structured according to the partitioning induced from $ \lambda_Q $ is equivalent, up to a normalization in the total derivative-energy of the signal, to the MSE of sampling the signal $ \varphi\left(t\right) $ according to the segmentation defined by $ \lambda_Q $. 
The MSE relation (\ref{eq:Bennet integral for quantization - relation to sampling MSE}) shows that the optimal $ \lambda_Q $ for the above quantization problem, is also optimal for sampling of the signal $ \varphi\left(t\right) $. Indeed, Bennett's integral for quantization of $ p_{\varphi}(x) $ is minimized for $ \lambda_Q $ formulated exactly as the optimal sampling-point density in (\ref{eq:sampling point density - one-dimensional signals - optimal - continuous}).

\subsubsection{Optimal Quantizer Design via Optimal Sampling}
~\\
Given a random variable $ X $ corresponding to the probability density function $ p_X (x) $ defined for $ x \in \left[x_{L},x_{H}\right) $, we can construct a signal $ \varphi_X \left(t\right) $ defined for $ t\in\left[0,1\right) $ via 
\begin{IEEEeqnarray}{rCl}
	\label{eq:equivalent signal to a given probability density function}
	\varphi_X \left(t \right) = \frac{1}{x_{H} - x_{L}}\mathop \int \limits_{x_{L}}^{x_{L} + t\cdot \left(x_{H} - x_{L}\right)} \sqrt{p_X (x)} dx .
\end{IEEEeqnarray}
We take the positive square-root of $p_X (x)$, hence, the signal $ \varphi_X \left(t \right) $ is monotonically non-decreasing. The derivative of $ \varphi_X \left(t \right) $ (with respect to $ t $) is 
\begin{IEEEeqnarray}{rCl}
	\label{eq:equivalent signal to a given probability density function - derivative energy}
	\varphi'_X \left(t \right) = \sqrt{p_X \big(x_{L} + t\cdot \left(x_{H} - x_{L}\right) \big)}  .
\end{IEEEeqnarray}
Then, according to (\ref{eq:Bennet integral for sampling - notations of quantization-sampling duality}), the sampling MSE of $ \varphi_X \left(t \right) $ for a sampling-point density $\lambda_S $ is 
\begin{IEEEeqnarray}{rCl}
	\label{eq:equivalent signal to a given probability density function - Bennet integral for sampling}
	\errcost_S \left(  \lambda_S ; \varphi _X \right) \approx \frac{1}{12 N^2} \mathop \int \limits_{0}^{1}  \frac{ \left({\varphi ' _X \left(t \right)}\right) ^2}{\lambda_S ^2 (t)} dt
\end{IEEEeqnarray}
that here equals to 
\begin{IEEEeqnarray}{rCl}
	\label{eq:equivalent signal to a given probability density function - Bennet integral for sampling - form 2}
	\errcost_S \left(  \lambda_S ; \varphi _X  \right) \approx \frac{1}{12 N^2} \mathop \int \limits_{0}^{1}  \frac{p_X \big( x_{L} + t\cdot \left(x_{H} - x_{L}\right) \big)}{\lambda_S ^2 (t)} dt .\nonumber\\
\end{IEEEeqnarray}
By changing the integration variable to $ x = x_{L} + t\cdot \left(x_{H} - x_{L}\right) $ we get 
\begin{IEEEeqnarray}{rCl}
	\label{eq:equivalent signal to a given probability density function - Bennet integral for sampling - form 3}
	\errcost_S \left(  \lambda_S ; \varphi _X  \right) \approx \frac{1}{12 N^2 \left(x_{H} - x_{L}\right)} \mathop \int \limits_{x_L}^{x_H}  \frac{p_X \left( x \right) }{\lambda_S ^2 (\frac{x - x_L }{x_{H} - x_{L}})} dx .\nonumber\\
\end{IEEEeqnarray}
By referring to (\ref{eq:Bennet integral for quantization}), the last form shows the following relation between the sampling MSE and the quantization MSE 
\begin{IEEEeqnarray}{rCl}
	\label{eq:equivalent signal to a given probability density function - Bennet integral for sampling - relation to quantization MSE}
	\errcost_S \left(  \lambda_S ; \varphi _X  \right) \approx \frac{1}{\left(x_{H} - x_{L}\right)} \errcost_Q \left(  \lambda_Q ; p_X  \right)
\end{IEEEeqnarray}
where, for $ x \in \left[x_{L},x_{H}\right) $,
\begin{IEEEeqnarray}{rCl}
	\label{eq:equivalent signal to a given probability density function - Bennet integral for sampling - relation to quantization MSE - point density relation}
	\lambda_Q  \left(  x \right) = \lambda_S \left( \frac{x - x_L }{x_{H} - x_{L}}  \right) .
\end{IEEEeqnarray}
The last result means that for a random variable $ X $, with a given probability density function $ p_X(x) $, one can design the optimal high-rate quantizer by considering the high-resolution sampling problem of the signal $ \varphi_X\left(t\right) $ defined in (\ref{eq:equivalent signal to a given probability density function}). Equations (\ref{eq:equivalent signal to a given probability density function - Bennet integral for sampling - relation to quantization MSE})-(\ref{eq:equivalent signal to a given probability density function - Bennet integral for sampling - relation to quantization MSE - point density relation}) show that the MSEs of the two procedures are equal up to a normalization by the width of the value-range of $ X $, and a linear transformation of the coordinates of the partitioning structure (see Eq. (\ref{eq:equivalent signal to a given probability density function - Bennet integral for sampling - relation to quantization MSE - point density relation})). Hence, implementing the optimal sampling-point density (\ref{eq:sampling point density - one-dimensional signals - optimal - continuous}) for the signal $ \varphi_X\left(t\right) $, together with the appropriate linear-transformation of the coordinates, provides the structure that minimizes the high-rate quantization MSE for $ p_X \left(x\right) $.

\subsection{Numerical Demonstrations}
\label{subsec:Numerical Demonstrations}

We now turn to study our theoretical results by applying them for analytic signals.
\subsubsection{Exponential Signals}
\begin{figure*}[]
	\centering
	{\subfloat[]{\label{fig:exponent__signal}\includegraphics[width=0.24\textwidth]{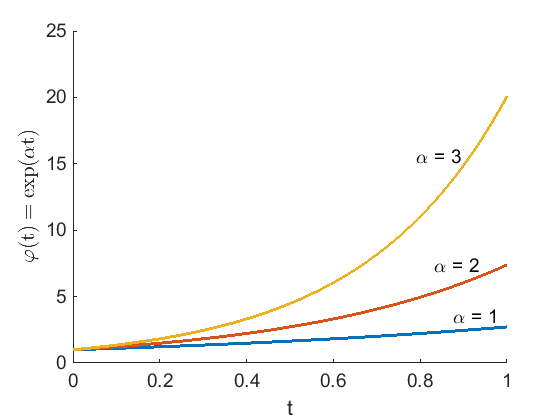}}}
	{\subfloat[]{\label{fig:exponent__signal_derivative_energy}\includegraphics[width=0.24\textwidth]{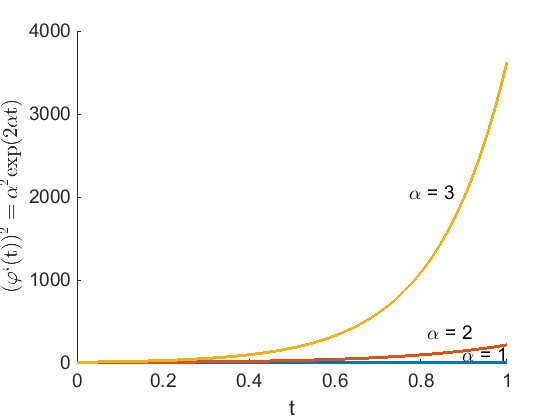}}}
	{\subfloat[]{\label{fig:exponent__error_for_alpha}\includegraphics[width=0.24\textwidth]{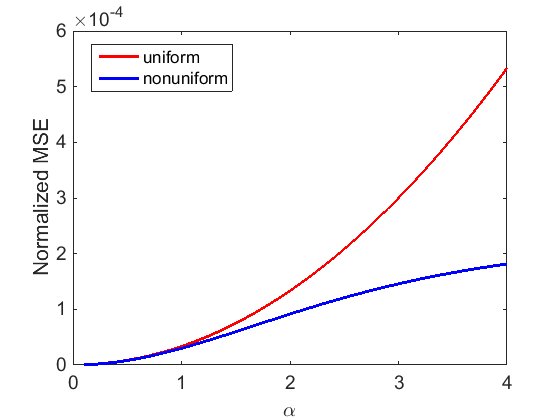}}}
	{\subfloat[]{\label{fig:exponent__error_for_N__alpha3}\includegraphics[width=0.24\textwidth]{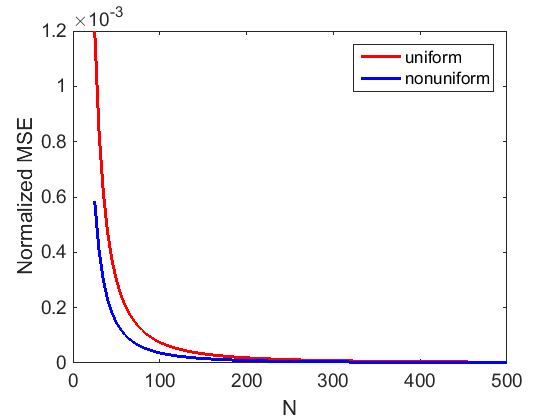}}}
	\caption{Demonstration for an exponential signal $ \varphi (t) = exp(\alpha t) $ for $ \alpha > 0 $. (a) The exponential signal for several $ \alpha $ values. (b) The signal-derivative energy for several $ \alpha $ values. In (c)-(d), theoretical reconstruction-MSE obtained via nonuniform and uniform sampling procedures are compared. (c) evaluated for various $ \alpha $ values and $ N=50 $. (d) evaluated for a range of sampling resolutions ($ N $) and $ \alpha = 3 $.} 
	\label{Fig:Exponent Signal Analytic Demonstration}
\end{figure*}

\begin{figure}[]
	\centering
	{\subfloat[Compressor Curve]{\label{fig:exponent_alpha3_compressor_curve}\includegraphics[width=0.36\textwidth]{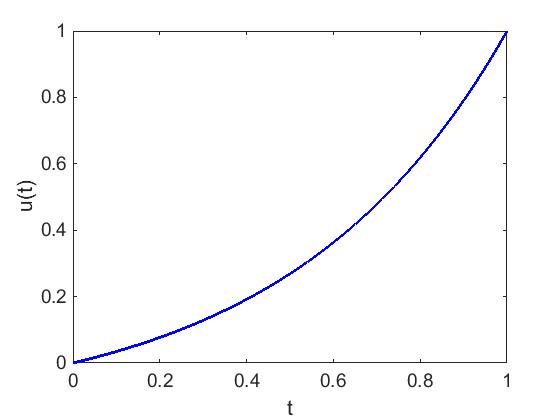}}}\\
	{\subfloat[Nonuniform Sampling]{\label{fig:exponent_alpha3_N50_nonuniform_sampling_reconstruction}\includegraphics[width=0.36\textwidth]{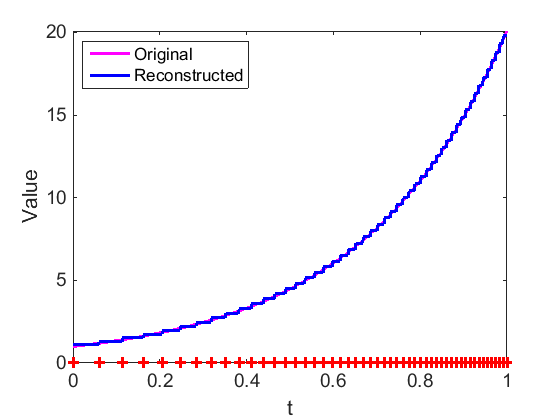}}}
	\caption{Optimal nonuniform sampling ($N=50$) of an exponential signal $ \varphi (t) = exp(3t) $. (a) the mapping between uniform to nonuniform sampling-spacing. (b) shows the original signal (magenta), the reconstructed signal from nonuniform sampling (blue), and the partitioning to sampling intervals (red). } 
	\label{Fig:Exponent Signal Analytic Demonstration - Example for alpha=3 and N=50}
\end{figure}

We consider an exponential signal of the form 
\begin{IEEEeqnarray}{rCl}
	\label{eq:exponential signal}
	\varphi (t) = e^{\alpha t}~~~,~t\in[0,1)
\end{IEEEeqnarray}
where $ \alpha > 0 $ is a real-valued parameter determining the growing rate (see examples in Fig. \ref{fig:exponent__signal}). The signal-derivative energy is expressed as $ ( \varphi ' (t) )^2 = \alpha ^2 e^{2\alpha t} $, which is positive valued for $ t\in[0,1) $ (Fig. \ref{fig:exponent__signal_derivative_energy}).
Therefore, the optimal sampling-point density for this monotonic signal is expressed as
\begin{IEEEeqnarray}{rCl}
	\label{eq:exponential signal - optimal sampling-point density}
	\lambda^{opt}\left(t\right) = \frac{2\alpha}{3}\cdot\frac{{e^{\frac{2}{3}\alpha t}} }{{e^{\frac{2}{3}\alpha}}-1 } 
\end{IEEEeqnarray}
Plugging (\ref{eq:exponential signal - optimal sampling-point density}) into (\ref{eq:optimal sampling expander function}) defines the optimal expander, $  v\left(\tau\right) $, via
\begin{IEEEeqnarray}{rCl}
	\label{eq:exponential signal - optimal expander}
 	\frac{{e^{\frac{2}{3}\alpha v\left(\tau\right)} }-1 }{ {e^{\frac{2}{3}\alpha}}-1 } = \tau
\end{IEEEeqnarray}
that yields
\begin{IEEEeqnarray}{rCl}
	\label{eq:exponential signal - optimal expander - explicit}
	v\left(\tau\right) = \frac{3}{2\alpha} \log \left( \left({e^{\frac{2}{3}\alpha } }-1\right)\cdot \tau + 1 \right), 
\end{IEEEeqnarray}
and the corresponding optimal nonuniform partitioning is determined via (\ref{eq:optimal sampling interval - boundaries - continuous}) as
\begin{IEEEeqnarray}{rCl}
	\label{eq:exponential signal - optimal boundaries}
	a_i^{opt} = \frac{3}{2\alpha} \log \left( \left({e^{\frac{2}{3}\alpha } }-1\right)\cdot \frac{i}{N} + 1 \right)  ~~~~,i=1,...,N.~
\end{IEEEeqnarray}
The MSE corresponding to the optimal nonuniform sampling is calculated using (\ref{eq:optimal sampling MSE}) and expressed as
\begin{IEEEeqnarray}{rCl}
	\label{eq:exponential signal - optimal sampling MSE}
	\errcost \left( {\left\{ {{a_i^{opt}}} \right\}_{i = 1}^{N - 1} }\right) = \frac{9}{32{\alpha} N^{2}} \left( e^{\frac{2}{3}\alpha} - 1 \right)^3.
\end{IEEEeqnarray}
We evaluate the gain of our approach with respect to the uniform sampling.  (for $ i=1,...,N $). The MSE of uniform high-resolution sampling is calculated by setting $a_i^{uniform} = \frac{i}{N}$ in the MSE expression in (\ref{eq:total sampling MSE - function of derivative}), yielding
\begin{IEEEeqnarray}{rCl}
	\label{eq:exponential signal - uniform sampling MSE}
	\errcost \left( {\left\{ {{a_i^{uniform}}} \right\}_{i = 1}^{N - 1} }\right) = \frac{\alpha}{24 N^2} \left( e^{2\alpha} - 1 \right).
\end{IEEEeqnarray}
The MSE of the two sampling methods are compared in Fig. \ref{fig:exponent__error_for_alpha} and Fig. \ref{fig:exponent__error_for_N__alpha3} for various exponential signals (via the $ \alpha $ parameter) and sampling resolutions (the parameter $ N $), respectively.
Note that the MSE values are normalized by the signal energy, here expressed as $ \int\limits_{t=0}^{1} \varphi ^2 (t) = \frac{1}{2\alpha} \left( e^{2\alpha} - 1 \right) $.
Figure \ref{Fig:Exponent Signal Analytic Demonstration - Example for alpha=3 and N=50} presents an example for the compressor function and the corresponding nonuniform sampling of an exponential signal.

\subsubsection{Cosine Signal}
\begin{figure*}[]
	\centering
	{\subfloat[]{\label{fig:cosine__alpha5_signal}\includegraphics[width=0.24\textwidth]{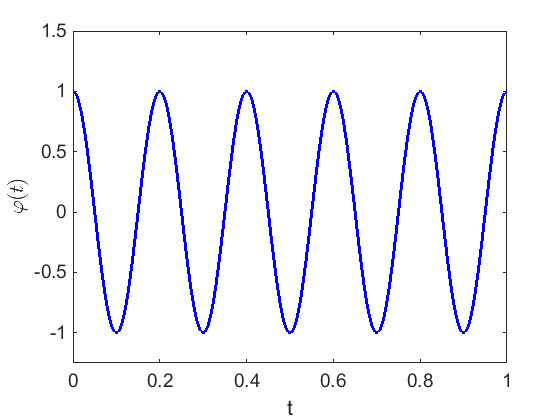}}}
	{\subfloat[]{\label{fig:cosine__alpha5_signal_derivative_energy}\includegraphics[width=0.24\textwidth]{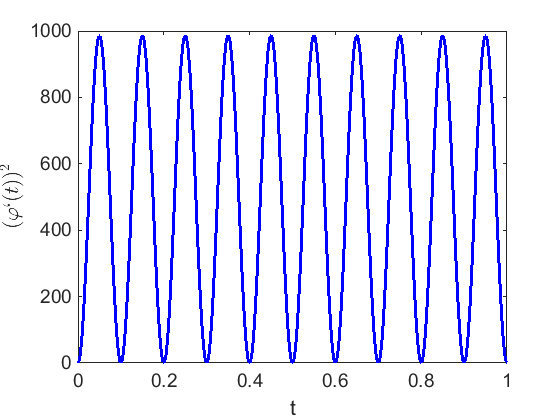}}}
	{\subfloat[]{\label{fig:cosine__alpha5_compressor_curve}\includegraphics[width=0.24\textwidth]{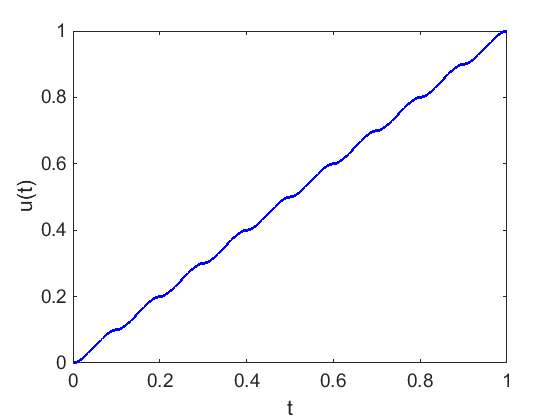}}}
	{\subfloat[]{\label{fig:cosine__alpha5_N100_nonuniform_sampling_reconstruction}\includegraphics[width=0.24\textwidth]{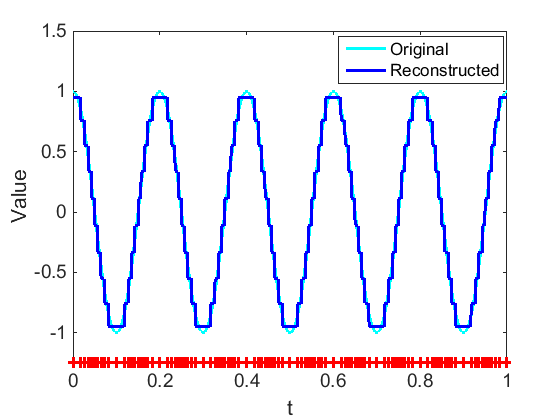}}}
	\caption{Demonstration for a cosine signal $ \varphi (t) = \cos(10\pi t) $. (a) The signal. (b) The signal-derivative energy. (c) The optimal compressor curve. (d) Optimal nonuniform sampling using $ N=100 $ samples (the partitioning of the signal domain is in red).} 
	\label{Fig:Cosine Signal Analytic Demonstration}
\end{figure*}

Let us demonstrate our approach for a non-monotonic signal of the form:
\begin{IEEEeqnarray}{rCl}
	\label{eq:cosine signal}
	\varphi (t) = \cos \left( 2\pi\alpha t \right) ~~~,~t\in[0,1)
\end{IEEEeqnarray}
where $ \alpha > 0 $ is an integer determining the number of periods contained in the $ [0,1) $ interval (see example for $ \alpha = 5 $ in Fig. \ref{fig:cosine__alpha5_signal}).
The derivative energy of the cosine signal in (\ref{eq:cosine signal}) is
\begin{IEEEeqnarray}{rCl}
	\label{eq:cosine signal - derivative energy}
	( \varphi ' (t) )^2  = 4 {\pi}^2 {\alpha}^2 \sin ^2 \left( 2\pi\alpha t \right) ~~~,~t\in[0,1) .
\end{IEEEeqnarray}
As demonstrated in Fig. \ref{fig:cosine__alpha5_signal_derivative_energy}, the signal-derivative energy is zero only at the points $ t=\frac{j}{2\alpha} $ for $ j=0,1,...,2\alpha-1 $.
The compressor function does not lend itself here to a simple analytic form, nevertheless, it can be constructed numerically via its definition as a cumulative-density-function (see Eq. (\ref{eq:optimal sampling compressor function})) providing the compressor curve in Fig. \ref{fig:cosine__alpha5_compressor_curve}. The $ 2\alpha $ points of zero signal-derivative energy are not a real obstacle in our numerical construction and, anyway, their corresponding values can be replaced by an arbitrarily small $ \varepsilon $ value\footnote{In the numerical demonstrations provided here we set the value of $ \varepsilon $ to the smallest positive floating-point value available in Matlab via the command \textit{eps}, which returns the value $ 2^{-52} $. } as suggested above.
The expander function is numerically formed as the inverse of the compressor curve. The resulting nonuniform sampling structure (Fig. \ref{fig:cosine__alpha5_N100_nonuniform_sampling_reconstruction}) shows its adaptation to the local signal derivative and to the periodic nature of the signal.

\subsubsection{Chirp Signal}
\begin{figure*}[]
	\centering
	{\subfloat[]{\label{fig:chirp__alpha5_signal}\includegraphics[width=0.24\textwidth]{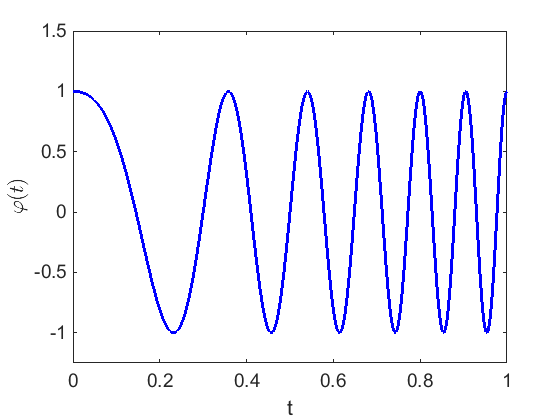}}}
	{\subfloat[]{\label{fig:chirp__alpha5_signal_derivative_energy}\includegraphics[width=0.24\textwidth]{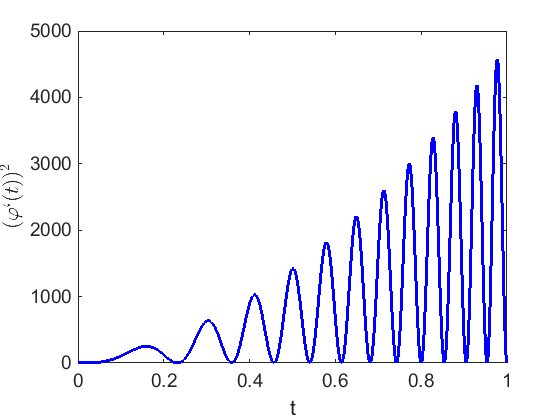}}}
	{\subfloat[]{\label{fig:chirp__alpha5_compressor_curve}\includegraphics[width=0.24\textwidth]{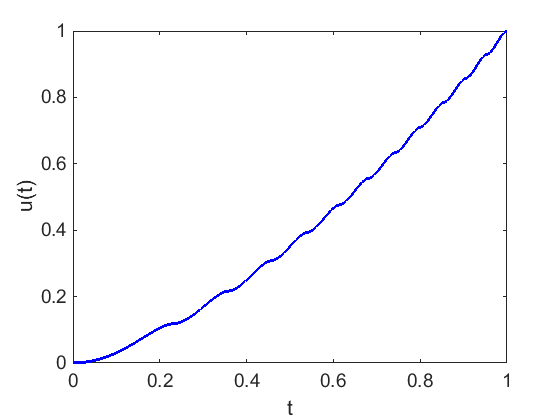}}}
	{\subfloat[]{\label{fig:chirp__alpha5_N100_nonuniform_sampling_reconstruction}\includegraphics[width=0.24\textwidth]{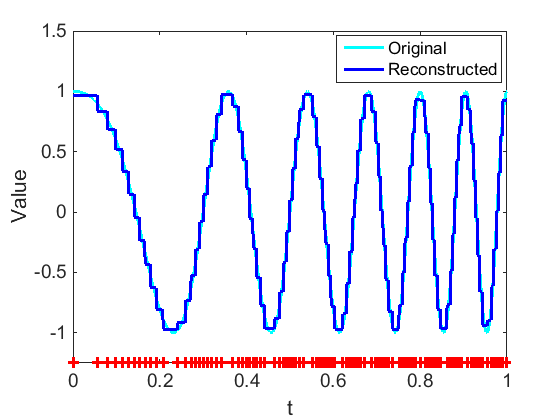}}}
	\caption{Demonstration for a chirp signal $ \varphi (t) = \cos\left(2\pi t \left( 1 + 5t \right) \right) $. (a) The signal. (b) The signal-derivative energy. (c) The optimal compressor curve. (d) Optimal nonuniform sampling using $ N=100 $ samples (the partitioning of the signal domain is in red).} 
	\label{Fig:Chirp Signal Analytic Demonstration}
\end{figure*}

The cosine signal in (\ref{eq:cosine signal}) can be extended to the following chirp signal with a linearly increasing frequency:
\begin{IEEEeqnarray}{rCl}
	\label{eq:chirp signal}
	\varphi (t) = \cos \left( 2\pi t \left(1+\alpha t\right) \right) ~~~,~t\in[0,1).
\end{IEEEeqnarray}
Here the $ \alpha > 0 $ parameter determines the linear growth-rate of the frequency (Fig. \ref{fig:chirp__alpha5_signal} exemplifies this for $ \alpha = 5 $).
The signal-derivative energy of the chirp (\ref{eq:chirp signal}) is 
\begin{IEEEeqnarray}{rCl}
	\label{eq:chirp signal - derivative energy}
	( \varphi ' (t) )^2  = 4 {\pi}^2 \left( 1 + 2 {\alpha} t \right)^2 \sin ^2 \left( 2\pi t \left( 1 +  \alpha t \right) \right) ~~~,~t\in[0,1) . \nonumber\\
\end{IEEEeqnarray}
The nonuniform sampling of the chirp is demonstrated in Fig. \ref{fig:chirp__alpha5_N100_nonuniform_sampling_reconstruction}. Comparison between the nonuniform sampling of the cosine signal (Fig. \ref{Fig:Cosine Signal Analytic Demonstration}) and the chirp signal (Fig. \ref{Fig:Chirp Signal Analytic Demonstration}) reveals the influence of the varying frequency embodied in the chirp.

\subsection{Experimental Evaluation of Nonuniform Segmentations}
\label{sec:Experimental Results - Sampling of One-Dimensional Signals}

In this section we present experimental evaluations of the procedure proposed in section \ref{sec:Analysis for One-Dimensional Signals} for nonuniform sampling of one-dimensional signals.

The proposed sampling method is compared to two other sampling approaches. The first is the trivial, however commonly-used, uniform sampling, where the signal domain is partitioned into equal-size sampling-intervals. Specifically, for a budget of $ N $ samples the signal domain $ [0,1) $ is segmented according to $ a_i = \frac{i}{N} $ for $ i=0,1,...,N $. The samples are determined as the averages of the corresponding sampling intervals.
The second competing method is a nonuniform sampling based on a binary-tree structure that is adapted to the signal via a Lagrange optimization, as presented in Appendix \ref{appendix:The Main Competing Method: Optimized Tree-Structured Nonuniform Sampling}.

\begin{figure}[]
	\centering
	{\subfloat[Cosine Signal]{\label{fig:experiments_cosine_curve_comparison}\includegraphics[width=0.24\textwidth]{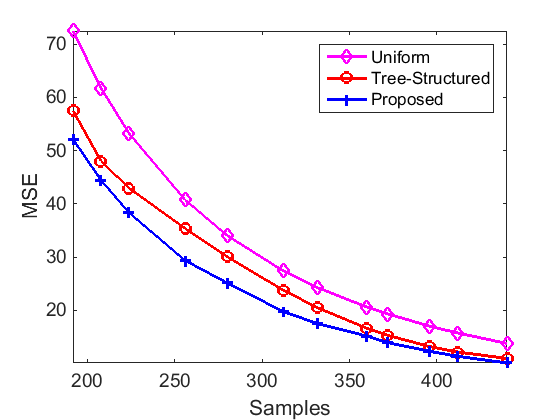}}}~~
	{\subfloat[Chirp Signal]{\label{fig:experiments_chirp_curve_comparison}\includegraphics[width=0.24\textwidth]{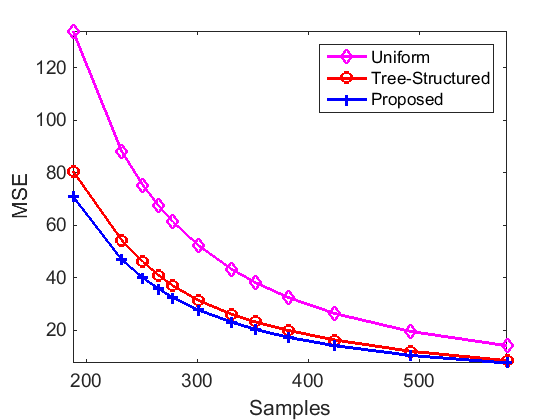}}}~~
	\caption{Performance comparison of the proposed sampling method, the
		uniform sampling approach, and the optimized tree-structured sampling. The curves present the sampling MSE obtained for various sample budgets.} 
	\label{Fig:Experiments - comparison of MSE-samples curves for several signals.}
\end{figure}

We examined sampling of several signals defined analytically. A grid of Lagrange multiplier values was set for the tree-structured Lagrange optimization (see Appendix \ref{appendix:The Main Competing Method: Optimized Tree-Structured Nonuniform Sampling}), this determined the number of samples to consider\footnote{Note that the proposed method and the uniform sampling do not rely on a Lagrange multiplier and operate directly based on a given number of samples, however, we defined the examined sample budgets based on the Lagrange multiplier grid of the tree-structured sampling in order to maintain an accurate comparison between all the sampling methods.}.
First, the sampling of the cosine signal $ \varphi\left(t\right) \nolinebreak=\nolinebreak 255\cdot \cos\left( 10\pi t\right) $ was examined, and showed that our method consistently outperforms the uniform and the tree-structured techniques for various amounts of samples (Fig. \ref{fig:experiments_cosine_curve_comparison}). These observations were further established by examining sampling of the chirp signal, $ \varphi\left(t\right) = 255\cdot \cos\left( 2\pi t \left(1+5t\right) \right) $ (Fig. \ref{fig:experiments_chirp_curve_comparison}).

\section{Practical Sampling Method for One-Dimensional Signals}
\label{sec:Practical Sampling Method}

In the former section we analytically studied the nonuniform sampling problem and formulated the corresponding optimal high-resolution segmentation. In this section we exhibit how the optimal nonuniform sampling design established above for continuous-time signals can be practically employed for nonuniform resampling of given discrete-time signals acquired via very high-resolution uniform sampling. Note that we refer to the practical resampling process using the term sampling, as it is indeed a sampling of a discrete signal that was already acquired in an initial high-resolution and uniformly sampled form.

Another issue of practical concern is that our unstructured partitioning of the time domain $ \left[0,1 \right) $ using the time-points $ \left\lbrace a_0^{opt} = 0, a_1^{opt}, ...,a_{N-1}^{opt},a_N^{opt} = 1 \right\rbrace $ means that a naive implementation will require the $ N-1 $ time points $ \left\lbrace a_i^{opt} \right\rbrace_{i=1}^{N-1} $ together with the $ N $ representation values $ \left\lbrace \varphi _i ^{opt} \right\rbrace_{i=1}^N $. This description cost of $ 2N-1 $ real values implies that a direct realization of the guidelines presented in Section \ref{sec:Analysis for One-Dimensional Signals} will result in an inefficient sampling process in terms of the performance measured by reconstruction error at a given representation cost. 

In this section we propose sampling and reconstruction procedures that employ the optimal segmentation found in Section \ref{sec:Analysis for One-Dimensional Signals} while requiring a significantly lower description cost than the naive approach. We show that a good approximation of the optimal reconstruction is possible by using only the segmentation time points, the signal extrema times and amplitude values, and some additional general properties of the signal. We also present experiments for compression of one-dimensional signals, showing that our approach outperforms the alternative of using an optimized adaptive tree-based segmentation.

\subsection{Practical Use of the Analytic Framework via Numerical Discretization}
\label{subsec:Practical Use of the Analytic Framework via Numerical Discretization}

The practical utilization of the proposed approach considers the nonuniform resampling system depicted in Fig. \ref{fig:general_sampling_reconstruction_procedure__practice}. An unknown continuous-time signal $ \varphi\left( t \right) $, defined for $ t\in\left[ 0,1\right) $, is uniformly sampled into a densely discrete representation using the $ N_U $ samples $ \left\lbrace \varphi\left(\frac{i}{N_U}\right)\right\rbrace_{i=0}^{i= N_U -1} $ that correspond to the fixed sampling interval of $ \Delta_U = \frac{1}{N_U} $. Then, this sequence of uniformly sampled values is resampled based on its nonuniform segmentation into $ N $ sub-sequences, each represented using a single sample that is later used for a piecewise-constant reconstruction of the signal. This resampling process is the discrete counterpart of the continuous-time problem studied in the previous section. Indeed, one can show that the optimal value representing a sub-sequence is its average value (here discretely computed via summation), just as the principle outlined in (\ref{eq:optimal sample}) for the continuous-time case. Moreover, we argue that the sampling framework developed in Section \ref{sec:Analysis for One-Dimensional Signals} for the continuous-time settings can be practically employed, based on numerical discretizations, for an efficient resampling of inputs corresponding to a high-resolution discrete-time uniform grid. This signal-processing concept of discrete implementation of ideas developed in a continuous-time framework is common, for example, in the well-established total-variation approach (see, e.g., \cite{rudin1992nonlinear}). Specifically, here we numerically approximate derivations and integrations using normalized-differences and summations. 
The reader should bear in mind the straightforward numerical discretizations needed in practice. However, for ease of presentation and connectivity to Section \ref{sec:Analysis for One-Dimensional Signals}, we present here our approach using continuous-time notions.

\subsection{Inferring Samples from a Given Optimal High-Resolution Segmentation}
\label{subsec:Inferring samples from segmentation}

The efficient utilization of the proposed optimal segmentation emerges from the following property and its implications. Notice that by Eq. (\ref{eq:optimal sampling interval - boundaries - continuous - explicit - sequential}) we can state that all the segments in the optimal partitioning contain an equal amount of the cube-root of the signal-derivative energy, this quantity is also defined in (\ref{eq:optimal sampling interval - sequential - optimal threshold}) as the threshold $ T_{opt} $ determining the segment lengths for the given signal and sample budget. Accordingly, we can merge  (\ref{eq:optimal sampling interval - boundaries - continuous - explicit - sequential})-(\ref{eq:optimal sampling interval - sequential - optimal threshold}) into the equation 
\begin{IEEEeqnarray}{rCl}
	\label{eq:practical sampling - segment property}
	{\int\limits_{a_{i-1}^{opt}}^{a_i^{opt}} {\sqrt[3]{ \left({\varphi ' \left(t \right)}\right) ^2   } dt}} = T_{opt}
\end{IEEEeqnarray}
obeyed for any segment $ \left[ a_{i-1}^{opt}, a_{i}^{opt} \right) $ for $ i=1,...,N $.
Recall that by our high-resolution assumption (presented in Eq. (\ref{eq:first-order Taylor approximation})) the sampling intervals are small enough such that the signal is well-approximated using a linear form, meaning also that the derivative is constant within each segment, i.e., 
\begin{IEEEeqnarray}{rCl}
	\label{eq:practical sampling - constant derivative within segment}
	{\varphi ' \left(t \right)} = {\varphi ' \left(t_i \right)}  \text{~~,  for ~} t\in\left[ a_{i-1}^{opt}, a_{i}^{opt} \right)
\end{IEEEeqnarray}
where $ \centeri $ is the center of the respective segment. Setting (\ref{eq:practical sampling - constant derivative within segment}) in (\ref{eq:practical sampling - segment property}) gives 
\begin{IEEEeqnarray}{rCl}
	\label{eq:practical sampling - segment property - constant derivative}
	\Delta_i^{opt} \sqrt[3]{ \left( {\varphi ' \left(\centeri \right)} \right) ^2   }  = T_{opt}
\end{IEEEeqnarray}
where $ \Delta_i^{opt} = {a_i^{opt}} - {a_{i-1}^{opt}} $ is the interval length.

Assume one knows the threshold $ T_{opt} $, the segment defining times $ a_{i-1}^{opt} $ and $ a_{i}^{opt} $, and whether the signal is monotonically increasing or decreasing in the considered interval (recall that by the high-resolution assumption the signal is locally linear and, thus, monotonic). Then, the relation (\ref{eq:practical sampling - segment property - constant derivative}) can be utilized for computing the local signal-derivative via 
\begin{IEEEeqnarray}{rCl}
	\label{eq:practical sampling - segment property - computing local derivative}
	{\varphi ' \left(t_i \right)} = s \cdot \sqrt{ \left( \frac{ T_{opt} } {\Delta_i^{opt} } \right)^{3} }
\end{IEEEeqnarray}
where $ s $ reflects the signal monotonicity in the segment by setting the derivative sign: if the signal is locally monotonic-decreasing then $ s = -1$, otherwise $ s = 1$.

Now, further assume that an estimate of the signal value at the beginning of the segment, $ \varphi _{est} \left(a_{i-1}\right) \approx \varphi \left(a_{i-1}\right) $, is also known. Hence, using the local-derivative obtained from (\ref{eq:practical sampling - segment property - computing local derivative}), we can form a linear estimate of the signal in the segment as 
\begin{IEEEeqnarray}{rCl}
	\label{eq:practical sampling - segment property - local linear approximation}
	\varphi _{est} (t) = \varphi _{est} \left(a_{i-1}\right) + \varphi ' \left(\centeri\right) \cdot \left(t- a_{i-1} \right) .
\end{IEEEeqnarray}
Consequently, using (\ref{eq:practical sampling - segment property - local linear approximation}) we can approximate the signal value at the segment center $\varphi _{est} (\centeri)$, that is also the estimate of the interval's average signal-value and, in our high-resolution case, is also the optimal sample $ \varphi_i^{opt} \approx \nolinebreak \varphi _{est} (\centeri) $ as appear in (\ref{eq:optimal sample - approximation following linearization}).

The developments above imply that a monotonic signal can be reconstructed using only the segmentation time points, the signal value at $ t=0 $, the monotonicity type (increasing or decreasing), and the threshold value $T_{opt}$. The process sequentially reconstructs the intervals by their order, for each interval the linear estimate (\ref{eq:practical sampling - segment property - local linear approximation}) is formed and used for two purposes: first, evaluating the optimal sample $\varphi_i^{opt}$ for the piecewise-constant approximation; and second, computing the signal value $ \varphi (a_i) $ for utilization in the reconstruction of the next segment, where it is the starting signal value. 
Since our goal is sampling of arbitrary non-monotonic signals, we present in the next subsection the required generalization of the above procedure.

\subsection{Sampling Method for Non-Monotonic Signals}
\label{subsec:Sampling Method for Non-Monotonic Signals}

The previous subsection motivates us to design a reconstruction procedure that sequentially goes over the segments and processes each based on three steps: inferring the local signal-derivative from the segment length, estimating the original signal segment using a linear model relying on the local derivative, and utilizing this estimate to approximate the optimal sample via averaging.
Using the local derivative for estimating the original signal segment requires knowing the local monotonicity of the signal, e.g., for setting the derivative sign in (\ref{eq:practical sampling - segment property - computing local derivative}). Commonly, the signal monotonicity changes in a finite number of signal extrema, located within segments set by the companding-based approach of Section \ref{sec:Analysis for One-Dimensional Signals}. Obviously, a segment containing one or more extrema cannot be well approximated using a linear form. Such an inadequate approximation of a segment may significantly affect the following intervals estimates due to the sequential dependency in the segment reconstructions. We address this issue as explained next.

Consider a signal with $ J $ extrema, occurring in times $ \left\lbrace x_j  \right\rbrace _{j=1}^{J} $ and having the corresponding signal values $ \left\lbrace \varphi_{x_{j}}  \right\rbrace _{j=1}^{J} $. We argue that in our high-resolution settings the number of samples is much larger than the number of extrema. Accordingly, we suggest to include the extrema times and amplitude values in the coded description produced by the sampler. Then, the reconstruction process can easily track the local monotonicity of the signal and also to better approximate the segments containing extrema, as will be explained next. 

The proposed reconstruction process gets the following data from the sampler: the $ N-1 $ segmentation times $ \left\lbrace a_i^{opt} \right\rbrace_{i=1}^{N-1} $, the $ J $ extrema descriptions $ \left\lbrace x_j, \varphi_{x_{j}} \right\rbrace _{j=1}^{J} $, the signal value at $ t = 0 $, the initial monotonicity of the signal (i.e., $ s_1 $), and the threshold value $ T_{opt} $. This yields a total description cost of $ N + 2J + 2 $ real values (in subsection ?? we will consider the task of compression, where the representation cost is expressed in bits). 
The suggested reconstruction is done via a sequential procedure reconstructing the segments. The reconstruction of a segment depends on whether it includes one or more of the (known) signal extrema: A segment with no extrema will be reconstructed as described in subsection \ref{subsec:Inferring samples from segmentation} using the derivative computation (\ref{eq:practical sampling - segment property - computing local derivative}) and the linear estimate (\ref{eq:practical sampling - segment property - local linear approximation}) that determines the sample for the piecewise-constant approximation; otherwise, the segment reconstruction will rely on the developments presented next.

Let us consider the reconstruction of a segment $ \left[ a_{i-1}, a_i \right) $ that contains a single extremum of value $ \varphi_{x_{j}} $ at $ t = x_j $.  We suggest to process such a segment based on its two sub-intervals: $  \left[ a_{i-1}, x_j \right) $ and $  \left[ x_j, a_i \right) $ where in each the signal is clearly monotonic of an opposing type. 
The segment information given to the reconstruction process allows forming a quadratic estimate for each of the sub-intervals.
We will use this nonlinear estimate for our original purpose of reconstructing the sample used in the piecewise-constant approximation, and also for approximating the signal value at $ t=a_i $ to be used in the next segment reconstruction. 

We start with the sub-interval  $  \left[ a_{i-1}, x_j \right) $, where the signal is to be estimated using the quadratic form of 
\begin{IEEEeqnarray}{rCl}
	\label{eq:practical sampling - segment with peak property - local quadratic approximation}
	\varphi_{est} (t) = \theta_2 t^2 + \theta_1 t + \theta_0  \text{~,~~for~} t\in  \left[ a_{i-1}, x_j \right] 
\end{IEEEeqnarray}
where $ \theta_0 , \theta_1 , $ and $ \theta _2 $ are real-valued parameters, determined using the following set of three linear equations. Note that the validity of (\ref{eq:practical sampling - segment with peak property - local quadratic approximation}) at $ t = x_j $ is for assuring the continuity of $ \varphi_{est} (t) $ in the sub-interval transition.
As for all segments, we have an estimate to the signal value at the interval start and, by (\ref{eq:practical sampling - segment with peak property - local quadratic approximation}), we can write the corresponding demand as 
\begin{IEEEeqnarray}{rCl}
	\label{eq:practical sampling - segment with peak property - local quadratic approximation - condition 1 - interval start}
	\theta_2 a_{i-1}^2 + \theta_1 a_{i-1} + \theta_0 = \varphi_{est} \left(a_{i-1}\right) . 
\end{IEEEeqnarray}
We also have the minimum/maximum value of $ \varphi_{x_{j}} $ at $ t = x_j $, implying that the estimate (\ref{eq:practical sampling - segment with peak property - local quadratic approximation}) value at $ t=x_j $ should obey 
\begin{IEEEeqnarray}{rCl}
	\label{eq:practical sampling - segment with peak property - local quadratic approximation - condition 2 - peak value}
	\theta_2 x_j^2 + \theta_1 x_j + \theta_0 = \varphi_{x_{j}} , 
\end{IEEEeqnarray}
and that the derivative of the estimate (\ref{eq:practical sampling - segment with peak property - local quadratic approximation}) at $ t=x_j $ is zero, i.e., 
\begin{IEEEeqnarray}{rCl}
	\label{eq:practical sampling - segment with peak property - local quadratic approximation - condition 3 - zero derivative at peak}
	2 \theta_2 x_j + \theta_1 = 0 . 
\end{IEEEeqnarray}
The solution of the linear equation set (\ref{eq:practical sampling - segment with peak property - local quadratic approximation - condition 1 - interval start})-(\ref{eq:practical sampling - segment with peak property - local quadratic approximation - condition 3 - zero derivative at peak}) provides the parameter values for the sub-interval estimate form in (\ref{eq:practical sampling - segment with peak property - local quadratic approximation}).

We continue with estimating the signal in the second sub-interval $  \left[ x_j, a_i \right) $, where the computation differs from the one of the former sub-interval. Again, we consider a quadratic signal estimate, formulated here as 
\begin{IEEEeqnarray}{rCl}
	\label{eq:practical sampling - segment with peak property - local quadratic approximation - subinterval 2}
	\varphi_{est} (t) = \rho_2 t^2 + \rho_1 t + \rho_0  \text{~,~~for~} t\in  \left[ x_j, a_i \right)
\end{IEEEeqnarray}
where $ \rho_0 , \rho_1 , $ and $ \rho_2 $ are real-valued parameters, determined using the following set of three equations (here the third equation will be nonlinear). The first two equations demand that the estimate  (\ref{eq:practical sampling - segment with peak property - local quadratic approximation - subinterval 2}) will be equal to the extremal value at $ t=x_j $, namely, 
\begin{IEEEeqnarray}{rCl}
	\label{eq:practical sampling - segment with peak property - local quadratic approximation - subinterval 2 - condition 1 - peak value}
	\rho_2 x_j^2 + \rho_1 x_j + \rho_0 = \varphi_{x_{j}} , 
\end{IEEEeqnarray}
and that the estimate derivative at $ t = x_j $ will satisfy 
\begin{IEEEeqnarray}{rCl}
	\label{eq:practical sampling - segment with peak property - local quadratic approximation - subinterval 2 - condition 2 - zero derivative at peak}
	2 \rho_2 x_j + \rho_1 = 0 . 
\end{IEEEeqnarray}
The third equation originates in our fundamental segment property presented in (\ref{eq:practical sampling - segment property}) that, using the sub-intervals defined here for the signal estimate, implies 
\begin{IEEEeqnarray}{rCl}
	\label{eq:practical sampling - segment with peak property - segment property}
	 {\int\limits_{x_j}^{a_{i}^{opt}} {\sqrt[3]{ \left({\varphi ' _{est} \left(t \right)}\right) ^2   } dt}} = T_{opt} - {\int\limits_{a_{i-1}^{opt}}^{x_j} {\sqrt[3]{ \left({\varphi ' _{est} \left(t \right)}\right) ^2   } dt}} . ~~
\end{IEEEeqnarray}
Using (\ref{eq:practical sampling - segment with peak property - local quadratic approximation}) and (\ref{eq:practical sampling - segment with peak property - local quadratic approximation - subinterval 2}), the last expression becomes 
\begin{IEEEeqnarray}{rCl}
	\label{eq:practical sampling - segment with peak property - segment property - explicit}
	{\int\limits_{x_j}^{a_{i}^{opt}} {\sqrt[3]{ \left( 2 \rho_2 t + \rho_1 \right) ^2   } dt}} = T_{opt} - {\int\limits_{a_{i-1}^{opt}}^{x_j} {\sqrt[3]{ \left( 2 \theta_2 t + \theta_1 \right) ^2   } dt}}  ~~~~
\end{IEEEeqnarray}
where $ \rho_2 $ and $ \rho_1 $ are unknown variables, and $ \theta_2 $ and $ \theta_1  $ are parameters already computed in the former sub-interval.
The equation system of (\ref{eq:practical sampling - segment with peak property - local quadratic approximation - subinterval 2 - condition 1 - peak value}), (\ref{eq:practical sampling - segment with peak property - local quadratic approximation - subinterval 2 - condition 2 - zero derivative at peak}), and (\ref{eq:practical sampling - segment with peak property - segment property - explicit}) can be solved (for the variables $ \rho_2 $,  $ \rho_1 $ and $ \rho_0 $) as follows. 
The equation pair ((\ref{eq:practical sampling - segment with peak property - local quadratic approximation - subinterval 2 - condition 1 - peak value})-(\ref{eq:practical sampling - segment with peak property - local quadratic approximation - subinterval 2 - condition 2 - zero derivative at peak})) can be translated into 
\begin{IEEEeqnarray}{rCl}
	\label{eq:practical sampling - segment with peak property - local quadratic approximation - subinterval 2 - condition 1 and 2 - translated - rho2}
	\rho_2 = - \frac{\rho_1}{2 x_j}
\end{IEEEeqnarray}
\begin{IEEEeqnarray}{rCl}
	\label{eq:practical sampling - segment with peak property - local quadratic approximation - subinterval 2 - condition 1 and 2 - translated - rho0}
	\rho_0 = \varphi_{x_{j}} - \frac{x_j \rho_1}{2}
\end{IEEEeqnarray}
expression $ \rho_2 $ and $ \rho_0 $ in terms of $\rho_1$.
Setting (\ref{eq:practical sampling - segment with peak property - local quadratic approximation - subinterval 2 - condition 1 and 2 - translated - rho2}) and (\ref{eq:practical sampling - segment with peak property - local quadratic approximation - subinterval 2 - condition 1 and 2 - translated - rho0}) into (\ref{eq:practical sampling - segment with peak property - segment property - explicit}) yields the following expression for the absolute value of $ \rho_1 $
\begin{IEEEeqnarray}{rCl}
	\label{eq:practical sampling - segment with peak property - segment property - translated - rho1}
	\left\lvert \rho_1 \right\rvert = \left(  \frac{ T_{opt}  -  {\int\limits_{a_{i-1}^{opt}}^{x_j} {\sqrt[3]{ \left( 2 \theta_2 t + \theta_1 \right) ^2   } dt}} } {   {\int\limits_{x_j}^{a_{i}^{opt}} {\sqrt[3]{ \left( 1 - \frac{t}{x_j} \right) ^2  }}}}  \right)^{\frac{3}{2}}	
\end{IEEEeqnarray}
where all the parameters in the right side of the equation are known, hence, the value of $ \left\lvert \rho_1 \right\rvert $ is computable.
Eq. (\ref{eq:practical sampling - segment with peak property - local quadratic approximation - subinterval 2 - condition 1 and 2 - translated - rho2}) implies that $ \rho_1 $ and $ \rho_2 $ have opposing signs. The value of $ \rho_2 $ is positive or negative depending on whether the signal extreme value at $ t=x_j $ is a minimum or a maximum, respectively, as can be determined by the signal monotonicity in the former segment. For example, if the previous signal segment is monotonically increasing, then, the value at $ t=x_j $ is a maximum and, accordingly, $ \rho_2 $ is negative and $ \rho_1 $ is positive. The complementary case mirrors this description. 
These sign determination rules let us to evaluate $ \rho_0 $, $ \rho_1 $, and $ \rho_2 $.
 
We essentially showed how for a segment with an extreme point, one can form quadratic estimates for its two sub-intervals. Averaging this signal segment estimate will provide an approximation of the optimal sample representing this interval in the piecewise-constant reconstruction. Recall that the extreme value implies that the local monotonicity-type changes and, thus, the corresponding variable $ s $ should also be updated for a correct reconstruction of the next segment.

As the sample budget decreases, the sampling intervals naturally increase and, accordingly, a segment is more likely to include more than one signal extreme value. This case is identifiable as the reconstruction receives all the extreme points' descriptions. Accordingly, we suggest to address such a segment in a similar way to the described above. First, the reconstruction procedure will be done with respect to the two sub-intervals defined by the first extreme point in the segment (this can be motivated by the assumption that the interval is still quite small). Second, the local monotonicity-type variable $ s $ will be set to describe the signal behavior after the last extreme point of the segment. 

The proposed sampling and reconstruction procedures are summarized in Algorithms \ref{Algorithm:Proposed Sampling Method}-\ref{Algorithm:Proposed Reconstruction Method}. 
We assume that the total number of samples, $ N $, is pre-defined and known to the reconstruction process, thus does not to be conveyed. Note that even if $ N $ is unknown and needs to be transmitted, its additional cost is marginal and does not affect the evaluations exhibited in this paper.

\begin{algorithm}
	\caption{Proposed Sampling Method}
	\label{Algorithm:Proposed Sampling Method}
	\begin{algorithmic}[1]
		\State Inputs: A signal $ \varphi\left(t\right) $ defined for $ t\in \left[ 0,1 \right) $.
		\newline .~~~~~~~ Number of samples to use $ N $.
		
		\State Compute the optimal segments $ \left\lbrace a_i^{opt} \right\rbrace _{i=0}^{N}$ via (\ref{eq:optimal sampling interval - boundaries - continuous}) or (\ref{eq:optimal sampling interval - boundaries - continuous - explicit - sequential}).
		\State  Determine the signal extrema times, $ \left\lbrace x_j \right\rbrace _{i=1}^{J}$, and values, $ \left\lbrace \varphi_{x_j} \right\rbrace _{i=1}^{J}$ .
		\State  Set the initial signal-monotonicity type $ s_0 $.
		\State  Set $ \varphi_{t=0} $ to the signal value at $ t = 0 $.
		\State  Compute the threshold $T_{opt}$ defined in (\ref{eq:optimal sampling interval - sequential - optimal threshold}).		
		\State Outputs:\newline  $ \left\lbrace a_i^{opt} \right\rbrace _{i=1}^{N-1}$, $ J $, $ \left\lbrace x_j \right\rbrace _{i=1}^{J}$, $ \left\lbrace \varphi_{x_j} \right\rbrace _{i=1}^{J}$, $ s_0 $, $ \varphi_{t=0} $, and $T_{opt}$ .
	\end{algorithmic}
\end{algorithm}

\begin{algorithm}
	\caption{Proposed Reconstruction Method}
	\label{Algorithm:Proposed Reconstruction Method}
	\begin{algorithmic}[1]
		\State Inputs:  $ \left\lbrace a_i^{opt} \right\rbrace _{i=1}^{N-1}$, $ J $, $ \left\lbrace x_j \right\rbrace _{i=1}^{J}$, $ \left\lbrace \varphi_{x_j} \right\rbrace _{i=1}^{J}$, $ s_0 $, $ \varphi_{t=0} $, and $T_{opt}$.
		\State Initialize $ i = 1 $, $ s = s_0 $, $ a_0^{opt} = 0 $, $ a_N^{opt} = 1 $.
		\State $ \varphi_{est}\left( a_0^{opt} \right) = \varphi_{t=0} $.
		
		\For{ $i=1,...,N$ } 
		
		\If {$ x_j \notin \left[ a_{i-1}^{opt} , a_{i}^{opt} \right)  ~,\forall j=1,...,J$} 
				\State Set the interval center as $ \centeri = \left( a_{i-1}^{opt} + a_{i}^{opt} \right) / 2$.
				\State Compute the local derivative, $ \varphi ' \left( \centeri \right) $, via (\ref{eq:practical sampling - segment property - computing local derivative}).

				\State \begin{varwidth}[t]{\linewidth} Use $ \varphi_{est}\left( a_{i-1}^{opt} \right) $ and $ \varphi ' \left( \centeri \right) $ to establish the  \par
					\hskip\algorithmicindent  linear estimate (\ref{eq:practical sampling - segment property - local linear approximation}) of the signal segment, \par
					\hskip\algorithmicindent  i.e., $ \varphi_{est}\left( t \right) $ for $ t \in \left[ a_{i-1}^{opt} , a_{i}^{opt}  \right] $.
				\end{varwidth}
				
				\State \begin{varwidth}[t]{\linewidth} Approximate the optimal sample as \par
					\hskip\algorithmicindent $ \varphi_i^{opt,est} = \varphi_{est} \left( \centeri \right) $ . \end{varwidth}

		\Else
				\State Set $ x_j^{first} = \min {x_j} \text{~~s.t.~~} x_j\in \left[ a_{i-1}^{opt} , a_{i}^{opt} \right)  $ .
				
				\State \begin{varwidth}[t]{\linewidth} Form the quadratic signal estimate, $ \varphi_{est}\left( t \right) $, for  \par
					\hskip\algorithmicindent the sub-interval $\left[ a_{i-1}^{opt} , x_j^{first} \right)$ as described in \par
					\hskip\algorithmicindent Eq. (\ref{eq:practical sampling - segment with peak property - local quadratic approximation})-(\ref{eq:practical sampling - segment with peak property - local quadratic approximation - condition 3 - zero derivative at peak}).  \end{varwidth}
				
				\State \begin{varwidth}[t]{\linewidth} Form the quadratic signal estimate, $ \varphi_{est}\left( t \right) $, for \par
					\hskip\algorithmicindent the sub-interval $\left[ x_j^{first} , a_{i}^{opt} \right]$ as described in \par
					\hskip\algorithmicindent Eq. (\ref{eq:practical sampling - segment with peak property - local quadratic approximation - subinterval 2},(\ref{eq:practical sampling - segment with peak property - local quadratic approximation - subinterval 2 - condition 1 and 2 - translated - rho2})-(\ref{eq:practical sampling - segment with peak property - segment property - translated - rho1}) and the related details.  \end{varwidth}

				\State \begin{varwidth}[t]{\linewidth} Approximate the optimal sample $ \varphi_i^{opt,est} $ as \par
					\hskip\algorithmicindent the average of $ \varphi_{est}\left( t \right) $ over $ \left[ a_{i-1}^{opt} , a_{i}^{opt} \right)  $. \end{varwidth}
				
				\State Set $ J_i  $ as the number of extrema in $ \left[ a_{i-1}^{opt} , a_{i}^{opt} \right)  $ .
				
				\State \begin{varwidth}[t]{\linewidth} Update the local monotonicity-type via \par
					\hskip\algorithmicindent $ s \leftarrow \left( -1 \right)^{J_i} s $ .  \end{varwidth}
		\EndIf
			
		\State \begin{varwidth}[t]{\linewidth} Set the segment in the piecewise-constant \par
			\hskip\algorithmicindent reconstructed signal as \par
			\hskip\algorithmicindent $ \hat{\varphi} \left( t \right) = \varphi_i^{opt,est} $ for $ t\in \left[ a_{i-1}^{opt} , a_{i}^{opt} \right) $ . \end{varwidth}
		
		\EndFor

		\State Output:  $ \hat{\varphi} \left( t \right) $ for $ t\in\left[ 0, 1 \right) $.
	\end{algorithmic}
\end{algorithm}

\subsection{Compression based on the Proposed Sampling Method}
\label{subsec:Compression based on the Proposed Sampling Method}

The sampling method presented in Algorithm \ref{Algorithm:Proposed Sampling Method} can be easily extended to a compression procedure for one-dimensional signals. This is done by encoding the sampler outputs, intended for the reconstruction process. We present here a specific compression implementation, relying on fixed-rate scalar quantizers and free of entropy coding. 

The segmentation times, $ \left\lbrace a_i^{opt} \right\rbrace _{i=1}^{N-1}$, and the extrema times, $ \left\lbrace x_j \right\rbrace _{i=1}^{J}$, are considered based on their discrete uniform high-resolution grid of the input signal that corresponds also to the resolution of the discrete approximation of time-continuous notions. Accordingly, we losslessly encode these sequences using a procedure designed for coding of monotonic sequences of integers (see Appendix \ref{appendix:Differential Coding of Monotonic Sequences of Integers}). 

The additional information required for the signal reconstruction is encoded as follows. 
The number of extrema, $ J $, is losslessly encoded using $ b_J $ bits, assuming that $ J \in \left\lbrace 0, ..., 2^{b_J}-1 \right\rbrace $).
The extreme values, $ \left\lbrace \varphi_{x_j} \right\rbrace _{i=1}^{J}$, are quantized using a uniform fixed-rate quantizer of $ b_{ext} $ bits considering the value range $ \left[ \varphi_{min}, \varphi_{max} \right] $. The starting monotonicity-type, $ s_0 $, can valued 1 or -1 and, thus, representable using a single bit. 
The signal value at $ t=0 $, $ \varphi_{t=0} $, is uniformly quantized using $ b_{val0} $ bits (the quantizer value range is $ \left[ \varphi_{min}, \varphi_{max} \right] $). 
The threshold $T_{opt}$ is encoded as a 64-bit floating-point value.

The reconstruction process remains the same as presented in Algorithm \ref{Algorithm:Proposed Reconstruction Method}, except for the need of decoding the input data. Of course that reconstruction quality is affected by the additional inaccuracies introduced in the quantization of  $ \left\lbrace \varphi_{x_j} \right\rbrace _{i=1}^{J}$ and $ \varphi_{t=0} $.

\subsection{Experimental Evaluation of the Proposed Method}
\label{subsec:Experimental Evaluation of the Proposed Method}

We conducted experiments to evaluate the practical sampling method presented in Algorithms \ref{Algorithm:Proposed Sampling Method}-\ref{Algorithm:Proposed Reconstruction Method} and its utilization for signal compression. The results provided here, for compression of analytic and audio signals, exhibit that our sampling method outperforms the optimized adaptive tree-based approach for large sample and bit budgets. 

Our compression implementation is as overviewed in Subsection \ref{subsec:Compression based on the Proposed Sampling Method}, with the specific parameters of $ b_J = 8 $,  $ b_{ext} = 13 $, $ \varphi_{min} = -255 $, $ \varphi_{max} = 255 $, and $ b_{val0} = 15 $. Recall that we do not need to encode the sample values.

We constructed a competing compression process by uniformly quantizing the samples provided by the optimized adaptive tree-based approach (a fixed rate of 8 bits per sample is used considering the value range of $ \left[ -255, 255 \right] $). Note that the quantization is not considered in the tree-structure optimization in order to maintain fairness with respect to our approach were the sampling is also independent of the quantization. The binary tree (representing the segmentation over the uniformly-spaced discrete grid of the input signal) is encoded into a sequence of bits using the procedure explained in \cite{shusterman1994image}.

\begin{figure}[]
	\centering
	{\subfloat[Cosine Signal]{\label{fig:experiments_compression_cosine_curve_comparison}\includegraphics[width=0.24\textwidth]{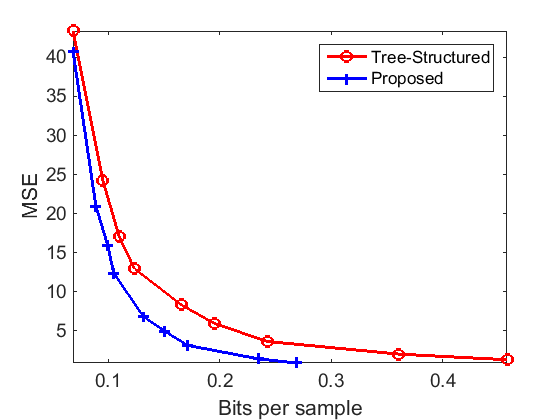}}}~~
	{\subfloat[Chirp Signal]{\label{fig:experiments_compression_chirp_curve_comparison}\includegraphics[width=0.24\textwidth]{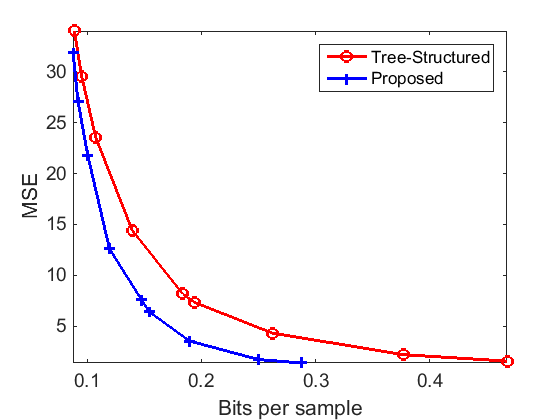}}}~~
	\caption{Performance comparison of compression based on the proposed sampling method and the optimized tree-structured sampling. The curves present the coding MSE obtained for various bit-rates (measured in bits per input sample, i.e., considering $ N_U $ and not $ N $).} 
	\label{Fig:Experiments - comparison of MSE-bpp curves for analytic signals}
\end{figure}

We consider again the cosine and chirp signals introduced in Fig. \ref{fig:cosine__alpha5_signal} and \ref{fig:chirp__alpha5_signal}, respectively, here scaled to have amplitude values in the range $ \left[ -255, 255 \right] $. The corresponding distortion-rate curves (Fig. \ref{Fig:Experiments - comparison of MSE-bpp curves for analytic signals}) exhibit that our method achieves more than 30\% bit-rate savings for the very high-resolution representations.

\begin{figure}[]
	\centering
	{\subfloat[An example of an audio signal]{\label{fig:Mozart_signal_full}\includegraphics[width=0.24\textwidth]{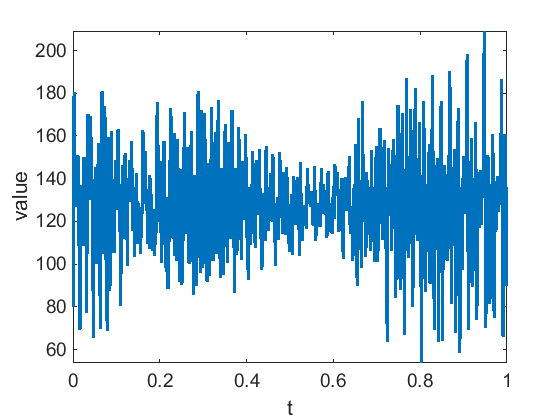}}}
	{\subfloat[Signal portion (Zoom-in)]{\label{fig:Mozart_signal_zoomin_segment}\includegraphics[width=0.24\textwidth]{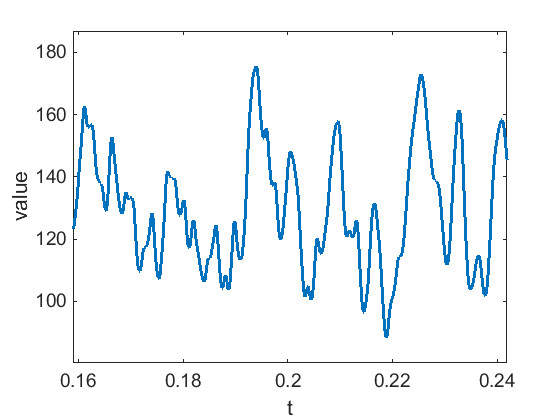}}}
	\caption{An exemplary audio signal ('Mozart') used in the experiments. (a) The complete signal considered. (b) Zoom-in for better view of the signal characteristics. The time axis of the signal is scaled to the interval $\left[ 0, 1\right)$ for compatibility to the settings of this paper. }
	\label{Fig:Experiments - exemplary real audio signal}
\end{figure}

\begin{figure}[]
	\centering
	{\subfloat['Mozart']{\label{fig:experiments_compression_real_Mozart_curve_comparison}\includegraphics[width=0.24\textwidth]{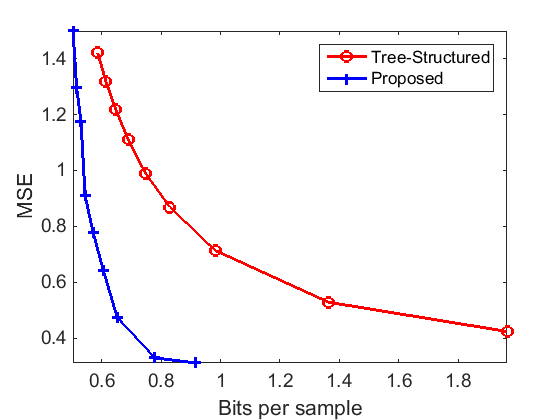}}}
	{\subfloat['Kleiberg']{\label{fig:experiments_compression_real_Kleiberg_curve_comparison}\includegraphics[width=0.24\textwidth]{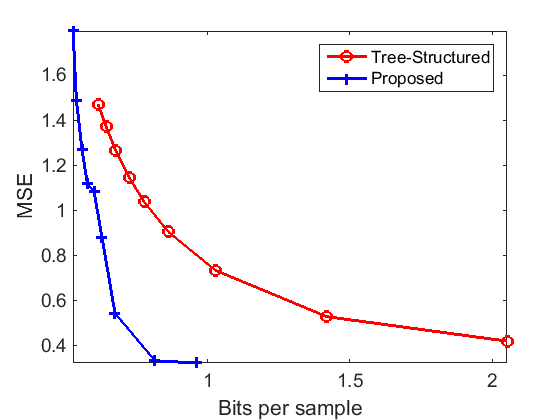}}}
	\caption{Performance comparison of compression based on the proposed sampling method and the optimized tree-structured sampling. The curves present the coding MSE obtained for various bit-rates (measured in bits per input sample, i.e., considering $ N_U $ and not $ N $).}
	\label{Fig:Experiments - comparison of MSE-bpp curves for several signals - audio}
\end{figure}

We evaluated our method also for compression of real (i.e., non-analytic) signals. For this purpose we used two audio signals (at an initial digital format of 24 bit per sample and sampling frequency of 192kHz) of classical music\footnote{The audio signals were downloaded from http://www.2l.no/hires/index.html.}, specifically, we considered segments of 65536 samples that were moderately smoothed using a Gaussian kernel (see an example for the obtained signal in Fig. \ref{Fig:Experiments - exemplary real audio signal}). The distortion-rate curves in Fig. \ref{Fig:Experiments - comparison of MSE-bpp curves for several signals - audio} show that, also for non-analytic signals, our method outperforms the use of optimized adaptive-tree segmentation.


\section{Analysis for Multidimensional Signals}
\label{sec:Analysis for Multidimensional Signals}
In this section we discuss the theoretic analysis of optimal high-resolution sampling by studying the problem for sampling multidimensional signals.
The analysis for the one-dimensional case is relatively simple since the partition is characterized by sampling-interval lengths (see section \ref{sec:Analysis for One-Dimensional Signals}). However, when considering the multidimensional problem the required analysis becomes more intricate as the sampling regions are, in general, of arbitrary shape and size. 
Accordingly, our analysis in this section is conceptually and mathematically similar to the study of multidimensional high-rate quantization provided by Gersho \cite{gersho1979asymptotically}, which generalized the one-dimensional theory of Bennett \cite{bennett1948spectra} and the distortion formula of Panter and Dite \cite{panter1951quantization}. 
Thus, we provide a theoretic framework for optimal multidimensional signal sampling at high resolution.

The given differentiable signal, $ \varphi \left(\vec{x}\right) $, is defined over a $K$-dimensional unit cube, $ \mathcal{C}^K \triangleq [0,1]^K $, and is scalar real-valued from a bounded range, i.e.,  $ \varphi \left(\vec{x}\right) \in \left[\varphi_L, \varphi_H \right] $ for any $ \vec{x} \in \mathcal{C}^K $.
The signal goes through a sampling procedure in order to provide a discrete representation using $ N\in\mathbb{N} $ (scalar valued) samples, $ \left\lbrace \varphi_i \right\rbrace _{i=1}^{N} $, corresponding to a partitioning of $ \mathcal{C}^K $ to $ N $ distinct multidimensional regions, $ \left\lbrace A_i \right\rbrace _{i=1}^{N} $, such that $ \cup_{i=1}^N A_i = \mathcal{C}^K $. 
Again, we consider a reconstruction procedure providing the continuous-domain piecewise-constant signal
\begin{IEEEeqnarray}{rCl}
	\label{eq:reconstructed continuous signal - multidimensional}
	\hat\varphi \left( \vec{x} \right) = {\varphi _i} \text{~~~~for~~~~}  \vec{x}\in A_i,
\end{IEEEeqnarray}
and optimization in the sense of minimizing the overall MSE, here formulated as (note the implicit normalization in the unit-cube volume)
\begin{IEEEeqnarray}{rCl}
	\label{eq:sampling MSE - multidimensional}
	\errcost \left( {\left\{ {{A_i}} \right\}_{i = 1}^{N},\left\{ {{\varphi _i}} \right\}_{i = 1}^{N}} \right) = \mathop \sum \limits_{i = 1}^N \mathop \int \limits_{A_i} {\left( {\varphi \left( \vec{x} \right) - \varphi _i} \right)^2}d\vec{x}. \nonumber\\
\end{IEEEeqnarray}
Moreover, as before, the optimal sampling coefficients given the partitioning $ \left\lbrace A_i \right\rbrace _{i=1}^{N} $ can be analytically determined, showing that
\begin{IEEEeqnarray}{rCl}
	\label{eq:optimal sample - multidimensional}
	{\varphi _i ^{opt}} = \frac{1}{V(A_i)} \mathop \int \limits_{A_i} \varphi \left( \vec{x} \right)d\vec{x},  ~~~~i=1,...,N 
\end{IEEEeqnarray}
where $ V(A_i) $ is the volume of the region $ A_i $.

We consider the case of optimal high-resolution sampling (i.e., $ N $ is very large), and assume that the optimal sampling regions are all small enough and appropriately shaped such that we can further presume that the signal $\varphi \left( \vec{x} \right)$ is well approximated within each region by a linear form that is locally determined. The last assumption emerges as a generalization of the assumptions for the one-dimensional case that were presented and analyzed in detail in Section \ref{subsec:Optimal High-Resolution Sampling}. Specifically, here  
\begin{IEEEeqnarray}{rCl}
	\label{eq:local linear approximation- multidimensional}
	\varphi (\vec{x}) \approx \vecslopeiTrans \vec{x} + \offseti  \text{~~~~for~~~~}  \vec{x}\in A_i,
\end{IEEEeqnarray}
where $ \vecslopei $ is a $ K $-dimensional column vector and $ \offseti $ is a scalar value.
Moreover, the above linear form can be determined in the $ i^{th} $ region by the first-order Taylor approximation of the signal around the region center, $ \veccenteri \triangleq \frac{1}{V(A_i)}  \mathop \int \limits_{A_i}  \vec{x} d\vec{x} $, namely,
\begin{IEEEeqnarray}{rCl}
	\label{eq:first-order Taylor approximation - multidimensional}
	\varphi (\vec{x}) \approx \varphi \left(\veccenteri\right) + {\signaljacobian{\vec{x}}{\veccenteri}}  \left(\vec{x}-\veccenteri\right)
\end{IEEEeqnarray}
where $ {\signaljacobian{\vec{x}}{\veccenteri}} $ is the signal gradient, evaluated at the region center, here having the form of a $ K $-length row vector consisting of the $ K $ partial derivatives, i.e.,
\begin{IEEEeqnarray}{rCl}
	\label{eq:multidimensional Jacobian matrix of the signal}
	\signaljacobian{\vec{x}}{\veccenteri} = \left[ {\signalPartialDerivative{x}{\veccenteri}{1}, \signalPartialDerivative{x}{\veccenteri}{2}, \ldots ,\signalPartialDerivative{x}{\veccenteri}{K}} \right], \nonumber\\ 
\end{IEEEeqnarray}
where $ \signalPartialDerivative{x}{\veccenteri}{j} $ is the partial derivative of the signal in the $ j^{th} $ standard direction ($ j=1,...,K $) measured at the region central point $ \veccenteri $.
Accordingly, the linear-form parameters of the $ i^{th} $ region are set to
\begin{IEEEeqnarray}{rCl}
	\label{eq:linear approximation parameters - multidimensional}
	\vecslopeiTrans &=& {\signaljacobian{\vec{x}}{\veccenteri}}
	\\
	\offseti &=& \varphi \left(\veccenteri\right) - {\signaljacobian{\vec{x}}{\veccenteri}} \veccenteri .
\end{IEEEeqnarray}

The local linear approximation within each region (\ref{eq:local linear approximation- multidimensional}) yields an optimal sampling coefficient that is approximately the signal value at the region center, i.e.,
\begin{IEEEeqnarray}{rCl}
	\label{eq:optimal sample - approximation following linearization - multidimensional}
	{\varphi _i ^{opt}} \approx \frac{1}{V(A_i)} \mathop \int \limits_{A_i} \left( \vecslopeiTrans \vec{x} + \offseti \right)d\vec{x} = \vecslopeiTrans \veccenteri + \offseti \approx \varphi \left(\veccenteri\right) . \nonumber \\
\end{IEEEeqnarray}
Then, the sampling MSE of the $ i^{th} $ region is
\begin{IEEEeqnarray}{rCl}
	\label{eq:sampling MSE of region - multidimensional}
	\errcost _i \left( A_{i} \right) & = & \frac{1}{V(A_i)} \mathop \int \limits_{A_i} {\left( {\varphi \left( \vec{x} \right) - \varphi _i^{opt}} \right)^2}d\vec{x}
	\\ \nonumber
	& \approx &  \frac{1}{V(A_i)}  \mathop \int \limits_{A_i} {\left[  {\vecslopeiTrans \vec{x} + \offseti - \left( \vecslopeiTrans \veccenteri + \offseti \right)    }\right]^2}d\vec{x}
	\\ \nonumber
	& = &  \frac{1}{V(A_i)} \norm{\vecslopei}_2^2 \mathop \int \limits_{A_i}  \norm {\vec{x} - \veccenteri }_2^2 d\vec{x}.
\end{IEEEeqnarray}

Inspired by the analysis given by Gersho \cite{gersho1979asymptotically} to high-rate quantization, we turn to interpret the last error expression using the normalized moment of inertia of the region $ A_i $ around its center $ \veccenteri $, defined as
\begin{IEEEeqnarray}{rCl}
	\label{eq:normalized moment of inertia of the i-th region}
	M(A_i) \triangleq \frac{\int \limits_{A_i} {\norm {\vec{x} - \veccenteri }_2^2} d\vec{x}}{K\cdot V(A_i)^{1+\frac{2}{K}} } ,
\end{IEEEeqnarray}
where this quantity is invariant to proportional scaling of the region. Now the region MSE becomes 
\begin{IEEEeqnarray}{rCl}
	\label{eq:sampling MSE of region - multidimensional - using moment}
	\errcost _i \left( A_{i} \right) & \approx & K \norm{\vecslopei}_2^2 M(A_i) V(A_i)^{\frac{2}{K}},
\end{IEEEeqnarray}
and the total MSE is expressed as
\begin{IEEEeqnarray}{rCl}
	\label{eq:sampling MSE - multidimensional - using moment}
	\errcost \left( \left\{ {{A_i}} \right\}_{i = 1}^{N} \right) = \mathop \sum \limits_{i = 1}^N V(A_i)\cdot \errcost _i \left( A_{i} \right)  \nonumber\\
	\approx K \mathop \sum \limits_{i = 1}^N \norm{\vecslopei}_2^2 M(A_i) V(A_i)^{1+\frac{2}{K}} .
\end{IEEEeqnarray}

We now assume that in the high-resolution scenario there is a sampling-point density function, $ \lambda\left(\vec{x}\right) $, such that for any small volume $A$ that contains $\vec{x}$, the fraction of sampling points contained in it is approximately  $ \lambda\left(\vec{x}\right) V\left(A\right)$. Furthermore, the density function satisfies the approximate relation
\begin{IEEEeqnarray}{rCl}
	\label{eq:sampling point density function}
	\lambda\left(\vec{x}\right) \approx \frac{1}{N\cdot V(A_i)}   ,  \text{~~~~for~~}  \vec{x}\in A_i .
\end{IEEEeqnarray}
Hence, the above assumption implies that adjacent sampling regions have similar density values.

Na and Neuhoff \cite{na1995bennett} introduced (in the context of vector quantization) the important notion of the inertial profile, denoted here as $ m\left(\vec{x}\right) $. The function $ m\left(\vec{x}\right) $ is assumed to be smooth  and to approximate the normalized moment of inertia  of the cells (around their mass centers) in the neighborhood of $ \vec{x} $. The smoothness of the inertial profile is based on the assumption that neighboring regions have similar values of normalized moment of inertia in a high-resolution segmentation.

The definitions of the sampling-point density function in (\ref{eq:sampling point density function}) and the inertial profile let us to express the MSE in (\ref{eq:sampling MSE - multidimensional - using moment}) as
\begin{IEEEeqnarray}{rCl}
	\label{eq:sampling MSE - multidimensional - sampling density}
	\errcost \left( \left\{ {{A_i}} \right\}_{i = 1}^{N} \right) \approx \frac{K}{N^{\frac{2}{K}}} \mathop \sum \limits_{i = 1}^N {\norm{\vecslopei}_2^2 \frac{m\left(\veccenteri\right)}{\lambda\left(\veccenteri\right)^{\frac{2}{K}}} V(A_i)} . \nonumber \\
\end{IEEEeqnarray}
Furthermore, due to the high-resolution assumption we approximate the previous sum by the following integral,
\begin{IEEEeqnarray}{rCl}
	\label{eq:sampling MSE - multidimensional - sampling density - integral - general with inertial profile}
	\errcost \left( \lambda, m \right) \approx \frac{K}{N^{\frac{2}{K}}} \mathop \int \limits_{\vec{x}\in{\mathcal{C}^K}} {\beta ^2 (\vec{x}) \frac{m\left(\vec{x}\right)}{\lambda\left(\vec{x}\right)^{\frac{2}{K}}} d\vec{x}} . \nonumber \\
\end{IEEEeqnarray}
where 
\begin{IEEEeqnarray}{rCl}
	\label{eq:continuous beta function - definition}
	{\beta ^2 (\vec{x}) } \triangleq \norm{\signaljacobian{\vec{z}}{\vec{x}}}_2^2 .
\end{IEEEeqnarray}
The error expression in (\ref{eq:sampling MSE - multidimensional - sampling density - integral - general with inertial profile}) can be interpreted as Bennett's integral for multidimensional sampling. This formula shows that the sampling MSE for a given signal, which is represented by its gradient-energy density, is determined by the sampling point density and the inertial profile of the sampling structure.

We now argue that optimal high-resolution sampling of a multidimensional linear signal with a uniform gradient-energy density evaluated as 1 everywhere in $\mathcal{C}^K$ is obtained by partitioning the signal domain,  $\mathcal{C}^K$, based on a tessellation generated by a single optimal (in the sense of minimum normalized moment of inertia) convex polytope, $A^*_K$, such that all the regions in the segmentation are congruent to it\footnote{Based on the high-resolution assumption, we neglect cells that are intersected by the boundary of the signal domain, $ \mathcal{C}^K $, and thus may not be congruent to the optimal tessellating polytope.}. Accordingly, this optimal division results in a constant normalized moment of inertia to all of the sampling regions, namely,
\begin{IEEEeqnarray}{rCl}
	\label{eq:optimal normalized moment of inertia}
	M(A_i) = M(A^*_K) \text{~~~~for~~~~} i=1,...,N ,
\end{IEEEeqnarray}
or in inertial profile terms: $ m\left(\vec{x}\right) = M(A^*_K) $, i.e., a constant function.
We further assume that optimal high-resolution sampling of a nonlinear signal is obtained by regions that are approximately congruent and scaled versions of the optimal polytope used for sampling a linear signal. Then, Eq. (\ref{eq:optimal normalized moment of inertia}) is satisfied also in our case of sampling a nonlinear signal, since scaling does not change the normalized moment of inertia.
The latter hypothesis mirrors the conjecture made by Gersho in \cite{gersho1979asymptotically} for high-rate quantization.
Consequently, the MSE expression (\ref{eq:sampling MSE - multidimensional - sampling density - integral - general with inertial profile}) is reformed to 
\begin{IEEEeqnarray}{rCl}
	\label{eq:sampling MSE - multidimensional - sampling density - integral}
	\errcost \left( \lambda \right) \approx \frac{K\cdot M(A^*_K)}{N^{\frac{2}{K}}} \mathop \int \limits_{\vec{x}\in{\mathcal{C}^K}} {\beta ^2 (\vec{x}) \frac{1}{\lambda\left(\vec{x}\right)^{\frac{2}{K}}} d\vec{x}} . \nonumber \\
\end{IEEEeqnarray}

We optimize the sampling procedure by characterizing the best sampling-point density, $ \lambda^{opt}(\vec{x}) $, that minimizes the MSE as expressed in (\ref{eq:sampling MSE - multidimensional - sampling density - integral}).
Similar to the optimization in the one-dimensional case (see Appendix \ref{appendix:Sampling-Point Density Optimization - one dimensional}), we rely on H\"{o}lder's inequality that provides us a lower bound to the MSE in (\ref{eq:sampling MSE - multidimensional - sampling density - integral}) in the form of 
\begin{IEEEeqnarray}{rCl}
	\label{eq:MSE lower bound from Holder}
	&& \errcost \left( \lambda \right)   \geq
	\\ \nonumber
	 	&& \frac{K\cdot M(A^*_K)}{N^{\frac{2}{K}}} \left( \mathop \int \limits_{~\vec{x}\in{\mathcal{C}^K}}{ \left( \beta ^2 (\vec{x}) \right)^{\frac{K}{K+2}} d\vec{x} } \right)^{1+\frac{2}{K}}
\end{IEEEeqnarray}
Here, following the application of H\"older's inequality, the MSE lower bound is attained when $ \lambda(\vec{x}) $ is proportional to $ \beta ^2 (\vec{x})\frac{1}{\left(\lambda(\vec{x})\right)^{\frac{2}{K}}} $ implying that the optimal sampling point density is
\begin{IEEEeqnarray}{rCl}
	\label{eq:optimal multidimensional sampling point density}
	\lambda^{opt}(\vec{x}) = \frac{\left( \beta ^2 (\vec{x}) \right)^{\frac{K}{K+2}}} {\mathop \int \limits_{\vec{z}\in{\mathcal{C}^K}}{ \left( \beta ^2 (\vec{z}) \right)^{\frac{K}{K+2}} d\vec{z} }} .
\end{IEEEeqnarray}
Note that we used the fact that integrating a density function over the entire domain should be 1. 

The optimal sampling-point density demonstrates that in regions where the derivative energy is higher, the sampling should be denser by reducing the volumes of the relevant sampling regions.
Moreover, returning to the discrete formulation for high-resolution sampling MSE in (\ref{eq:sampling MSE - multidimensional - using moment}) and by utilizing (\ref{eq:sampling point density function}) together with (\ref{eq:optimal multidimensional sampling point density}) and (\ref{eq:optimal normalized moment of inertia}) shows that in the optimal solution all the sampling regions contribute the same amount of MSE.
In addition, the results in this section are generalization of those obtained in the analysis of one-dimensional signals in the previous section, this can be observed by setting $ K=1 $ and $ M(A^*_1) = \frac{1}{12} $, which is the normalized moment of inertia for one-dimensional intervals (or any other $ K $-dimensional cube) around their center.

The high-resolution analysis provides a theoretic evaluation of a sampler based on its sampling-point density function. 
As we described above, the suggested framework lets to determine the optimal sampling procedure in terms of the best sampling-point density function.
In the one-dimensional case the sampling-point density can be directly translated to a practical sampling procedure via the companding model (see sections \ref{sec:Analysis for One-Dimensional Signals} and \ref{subsec:Experimental Results - Evaluation and Comparison}).
However, in the multidimensional case, in general, there are no direct ways to implement a sampler based on a given sampling-point density function. 
This conclusion is based on the following results from the quantization field \cite{gersho1979asymptotically,bucklew1981companding,bucklew1983note}.
In \cite{bucklew1981companding,bucklew1983note} it was shown that optimal companding requires a compressor function that is a conformal mapping. 
This result implies that a direct implementation of multidimensional sampler based on a given point density is limited to a minor class of signals with a suitable gradient-energy density -- thus, in general, optimal multidimensional companding is impractical. 
This result was followed by a treatment of multidimensional companding for the vector quantization problem in limited settings that consider suboptimal solutions and/or particular source distributions \cite{moo1997optimal,simon1998suboptimal,samuelsson2003multidimensional,peric2008optimal}. 
Consequently, as in the high-rate quantization literature, the analysis provided here for the multidimensional case is a theoretic framework for studying sampling of multidimensional signals. Specifically, it describes the optimal sampler and allows to assess its theoretic performance. This together with Bennett's integral for multidimensional sampling (Eq. (\ref{eq:sampling MSE - multidimensional - sampling density - integral - general with inertial profile})) can be used to evaluate the performance of practical suboptimal sampling procedures (similar to the analysis of practical vector quantizers in \cite{na1995bennett}).

\section{Conclusion}
\label{sec:Conclusion}
We analyze the topic of nonuniform sampling of deterministic signals as a mirror-image of nonuniform quantization.
With the advent of new technologies, adaptive sampling becomes a viable alternative to be considered in data compression applications.
In all the above developments, the crucial local-density controlling parameter turns out to be the local energy of the signal gradient.
A new adaptive sampling method for one-dimensional signals is proposed and experimentally established as a leading nonuniform-sampling approach.


%

\appendices 

\section{Analysis of Inaccuracies Due to the Signal Linearity Assumption}
\label{appendix:Analysis of Inaccuracies Due to the Signal Linearity Assumption}

Our main high-resolution assumption suggests to consider the signal via its local linear approximation. 
Let us assume that the second derivative of the signal, $ \varphi '' (t) $, is continuous and bounded, i.e.,  $ \left\lvert \varphi '' (t) \right\rvert \le M $ for some positive constant $ M $.
Expressing the remainder of the first-order Taylor approximation using its integral form, lets us to rewrite (\ref{eq:first-order Taylor approximation}), for $ t \in \left[a_{i-1},a_i\right)$ (recall that $ t_i $ is the interval center), as 
\begin{IEEEeqnarray}{rCl}
	\label{eq:first-order Taylor approximation - remainder in integral form}
	\varphi (t) = \varphi \left(\centeri\right) + \varphi ' \left(\centeri\right) \cdot \left(t-\centeri\right) + R_i(t) , \nonumber\\
\end{IEEEeqnarray}
where the remainder (for the $ i^{th} $ sampling interval) is 
\begin{IEEEeqnarray}{rCl}
	\label{eq:remainder in integral form}
	R_i(t) = \int\limits_{t_i}^{t} \left( t - z \right) \varphi '' \left(z\right) dz
\end{IEEEeqnarray}
Considering the remainder in its integral form will be useful for the analysis in this appendix. 
The absolute value of the remainder is bounded for $ t\in \left[a_{i-1},a_i\right) $ as follows 
\begin{IEEEeqnarray}{rCl}
	\label{eq:remainder in integral form - absolute value bound}
	\left\lvert R_i \left( t \right) \right\rvert \le \frac{M}{2} \left( t - t_i \right)^2
	  \nonumber\\
\end{IEEEeqnarray}
where the last inequality conforms with $ R_i \left( t \right) = o(\left\lvert t - t_i \right\rvert) $ for $ t \rightarrow t_i $, as in Eq. (\ref{eq:first-order Taylor approximation}).

As explained in Section \ref{sec:Analysis for One-Dimensional Signals}, the optimal sample value is the signal average over the sampling interval (see Eq. (\ref{eq:optimal sample})). Therefore, averaging the signal in its form from (\ref{eq:first-order Taylor approximation - remainder in integral form}) gives 
\begin{IEEEeqnarray}{rCl}
	\label{eq:optimal sample - with remainder in integral form}
	{\varphi _i ^{opt}} =  \varphi \left(\centeri\right) + \frac{1}{\Delta _i} \mathop \int \limits_{{a_{i - 1}}}^{{a_i}} R_i (t) dt .
\end{IEEEeqnarray}
Hence, the remainder average is the amount of inaccuracy in the optimal sample value due to the signal linearity assumption. We analyze this quantity by bounding it 
\begin{IEEEeqnarray}{rCl}
	\label{eq:optimal sample - inaccuracy bound}
	\left\lvert \frac{1}{\Delta _i} \mathop \int \limits_{{a_{i - 1}}}^{{a_i}} R_i (t) dt \right\rvert & \le &  \frac{1}{\Delta _i} \mathop \int \limits_{{a_{i - 1}}}^{{a_i}} \left\lvert R_i (t) \right\rvert dt \\ \nonumber
	& \le & \frac{M}{2\Delta _i} \mathop \int \limits_{{a_{i - 1}}}^{{a_i}} \left( t - t_i \right)^2  dt = \frac{M}{24} \Delta_i^2
\end{IEEEeqnarray}
where we used the remainder bound from (\ref{eq:remainder in integral form - absolute value bound}). Using the bound in (\ref{eq:optimal sample - inaccuracy bound}) we can state that the inaccuracy in the sample value is 
\begin{IEEEeqnarray}{rCl}
	\label{eq:optimal sample - inaccuracy - little o}
	\frac{1}{\Delta _i} \mathop \int \limits_{{a_{i - 1}}}^{{a_i}} R_i (t) dt = o(\Delta_i)
\end{IEEEeqnarray}
for $ \Delta_i \rightarrow 0 $, as was presented in (\ref{eq:optimal sample - approximation following linearization}).

Now we proceed to analyzing the sampling MSE in the $ i^{th} $ interval. The basic expression given in (\ref{eq:sampling MSE of subinterval }) is equivalent to 
\begin{IEEEeqnarray}{rCl}
	\label{eq:sampling MSE of subinterval - appendix form}
	MSE_i = \frac{1}{\Delta _i} \mathop \int \limits_{{a_{i - 1}}}^{{a_i}} \varphi ^2 (t) dt - \left( {\varphi _i ^{opt}} \right)^2
\end{IEEEeqnarray}
Then, using the expressions from (\ref{eq:first-order Taylor approximation - remainder in integral form}) and (\ref{eq:optimal sample - with remainder in integral form}) the interval MSE becomes 
\begin{IEEEeqnarray}{rCl}
	\label{eq:sampling MSE of subinterval - with inaccuracy term}
	MSE_i & = & \frac{1}{12} \left({\varphi ' \left(\centeri\right)}\right) ^2 {\Delta _i}^2  +  D_i
\end{IEEEeqnarray}
where 
\begin{IEEEeqnarray}{rCl}
	\label{eq:sampling MSE of subinterval - formula of inaccuracy using remainder}
	D_i & = & \frac{1}{\Delta _i} \mathop \int \limits_{{a_{i - 1}}}^{{a_i}} R_i^2 (t) dt \\\nonumber 
	&& + 2{\varphi ' \left(\centeri\right)} \frac{1}{\Delta _i} \mathop \int \limits_{{a_{i - 1}}}^{{a_i}} \left(t - t_i \right) R_i (t) dt \\ \nonumber 
	&& - \left( \frac{1}{\Delta _i} \mathop \int \limits_{{a_{i - 1}}}^{{a_i}} R_i (t) dt  \right)^2
\end{IEEEeqnarray}
evaluates the deviation from the MSE obtained for the linear-approximation signal. 
Let us bound the MSE deviation term, $ D_i $, as follows:  
\begin{IEEEeqnarray}{rCl}
	\label{eq:sampling MSE of subinterval - formula of inaccuracy using remainder - bound}
	&&\left\lvert D_i \right\rvert \le \\ \nonumber
	&& \le \frac{1}{\Delta _i} \mathop \int \limits_{{a_{i - 1}}}^{{a_i}} R_i^2 (t) dt + \frac{2}{\Delta _i} \left\lvert  {\varphi ' \left(\centeri\right)}  \right\rvert \left\lvert    \mathop \int \limits_{{a_{i - 1}}}^{{a_i}} \left(t - t_i \right) R_i (t) dt \right\rvert .
\end{IEEEeqnarray}
Then, using the bound (\ref{eq:remainder in integral form - absolute value bound}) we get 
\begin{IEEEeqnarray}{rCl}
	\label{eq:sampling MSE of subinterval - formula of inaccuracy using remainder - bound - first part}
	&& \frac{1}{\Delta _i} \mathop \int \limits_{{a_{i - 1}}}^{{a_i}} R_i^2 (t) dt \le \frac{M^2}{4\Delta _i} \mathop \int \limits_{{a_{i - 1}}}^{{a_i}} \left( t - t_i \right)^4 dt = \frac{M^2}{320} \Delta_i^4 , 
\end{IEEEeqnarray}
and also 
\begin{IEEEeqnarray}{rCl}
	\label{eq:sampling MSE of subinterval - formula of inaccuracy using remainder - bound - part of second part}
	\left\lvert    \mathop \int \limits_{{a_{i - 1}}}^{{a_i}} \left(t - t_i \right) R_i (t) dt \right\rvert & \le & \mathop \int \limits_{{a_{i - 1}}}^{{a_i}} \left\lvert t - t_i \right\rvert \left\lvert R_i (t) \right\rvert dt \\ \nonumber
	& \le & \frac{M}{2} \mathop \int \limits_{{a_{i - 1}}}^{{a_i}} \left\lvert t - t_i \right\rvert ^3 dt =  \frac{M}{64} \Delta_i^4 .
\end{IEEEeqnarray}
Plugging (\ref{eq:sampling MSE of subinterval - formula of inaccuracy using remainder - bound - first part}) and (\ref{eq:sampling MSE of subinterval - formula of inaccuracy using remainder - bound - part of second part}) into (\ref{eq:sampling MSE of subinterval - formula of inaccuracy using remainder - bound}), together with assuming that $ \left\lvert  {\varphi ' \left(\centeri\right)}  \right\rvert < M_1 $, yields 
\begin{IEEEeqnarray}{rCl}
	\label{eq:sampling MSE of subinterval - formula of inaccuracy using remainder - bound - 2}
	\left\lvert D_i \right\rvert \le  \frac{M^2}{320} \Delta_i^4 + \frac{M\cdot M_1}{32} \Delta_i^3
\end{IEEEeqnarray}
implying that the MSE deviation follows 
\begin{IEEEeqnarray}{rCl}
	\label{eq:sampling MSE of subinterval - formula of inaccuracy using remainder - bound - little o }
	D_i = o\left( \Delta_i^2 \right) 
\end{IEEEeqnarray}
as $ \Delta_i \rightarrow 0 $.

\section{Sampling-Point Density Optimization via H\"{o}lder's Inequality for The One-Dimensional Case}
\label{appendix:Sampling-Point Density Optimization - one dimensional}

We aim at minimizing the sampling MSE given in (\ref{eq:Bennet integral for sampling}) as 
\begin{IEEEeqnarray}{rCl}
	\label{eq:Bennet integral for sampling - appendix}
	\errcost \left(  \lambda  \right) \approx \frac{1}{12 N^2} \mathop \int \limits_{0}^{1}  \frac{ \left({\varphi ' \left(t \right)}\right) ^2}{\lambda ^2 (t)} dt.
\end{IEEEeqnarray}

Using H\"{o}lder's inequality we can write 
\begin{IEEEeqnarray}{rCl}
	\label{eq:Holder's inequality for one-dimensional optimization - appendix}
{\left( {\mathop \int \limits_{0}^{1} {{\left( {\sqrt[3]{{{{\left( {{\varphi '}\left( t \right)} \right)}^2}}} \cdot \frac{1}{{{\lambda ^{\frac{2}{3}}}\left( t \right)}}} \right)}^3}dt} \right)^{\frac{1}{3}}}{\left( {\mathop \int \limits_{0}^{1} {{\left( {{\lambda ^{\frac{2}{3}}}\left( t \right)} \right)}^{\frac{3}{2}}}dt} \right)^{\frac{2}{3}}} 
\nonumber \\ \nonumber
\geqslant \mathop \int \limits_{0}^{1} \sqrt[3]{{{{\left( {{\varphi '}\left( t \right)} \right)}^2}}}dt . \\
\end{IEEEeqnarray}
Simplifying the left side of (\ref{eq:Holder's inequality for one-dimensional optimization - appendix}) gives 
\begin{IEEEeqnarray}{rCl}
\label{eq:Holder's inequality for one-dimensional optimization - simplified - appendix}
{\left( {\mathop \int \limits_0^1 {{\left( {{\varphi '}\left( t \right)} \right)}^2} \cdot \frac{1}{{{\lambda ^2}\left( t \right)}}dt} \right)^{\frac{1}{3}}}{\left( {\mathop \int \limits_0^1 \lambda \left( t \right)dt} \right)^{\frac{2}{3}}} \\
\geqslant \mathop \int \limits_0^1 \sqrt[3]{{{{\left( {{\varphi '}\left( t \right)} \right)}^2}}}dt . \nonumber
\end{IEEEeqnarray}
Then, since $ \lambda\left(t\right) $ is a density function its integration over the entire domain equals to 1, reducing (\ref{eq:Holder's inequality for one-dimensional optimization - simplified - appendix}) into 
\begin{IEEEeqnarray}{rCl}
	\label{eq:Holder's inequality for one-dimensional optimization - simplified 2 - appendix}
	\mathop \int \limits_0^1 {\left( {{\varphi '}\left( t \right)} \right)^2} \cdot \frac{1}{{{\lambda ^2}\left( t \right)}}dt \geqslant {\left( {\mathop \int \limits_0^1 \sqrt[3]{{{{\left( {{\varphi '}\left( t \right)} \right)}^2}}}dt} \right)^3} .
\end{IEEEeqnarray}

Here, H\"{o}lder's inequality is attained with equality when $ \lambda\left(t\right) $ is proportional to $ {\left( {{\varphi '}\left( t \right)} \right)^2} \cdot \frac{1}{{{\lambda ^2}\left( t \right)}}$ , therefore, the optimal sampling-point density is  
\begin{IEEEeqnarray}{rCl}
\label{eq:optimal one-dimensional sampling-point density - appendix}
{\lambda ^{opt}}\left( t \right) = \frac{{\sqrt[3]{{{{\left( {{\varphi '}\left( t \right)} \right)}^2}}}}}{{\mathop \int \nolimits_0^1 \sqrt[3]{{{{\left( {{\varphi '}\left( z \right)} \right)}^2}}}dz}} .
\end{IEEEeqnarray}
Setting the lower bound from (\ref{eq:Holder's inequality for one-dimensional optimization - simplified 2 - appendix}), which is achieved by $ \lambda^{opt}(t) $, in the MSE expression from (\ref{eq:Bennet integral for sampling - appendix}) gives the following optimal sampling MSE 
\begin{IEEEeqnarray}{rCl}
\label{eq:optimal one-dimensional sampling MSE - appendix}
MSE\left( {{\lambda ^{opt}}} \right) \approx \frac{1}{{12{N^2}}}{\left( {\mathop \int \limits_0^1 \sqrt[3]{{{{\left( {{\varphi '}\left( t \right)} \right)}^2}}}dt} \right)^3} .
\end{IEEEeqnarray}

\section{The Main Competing Sampling Method: Tree-Structured Nonuniform Sampling}
\label{appendix:The Main Competing Method: Optimized Tree-Structured Nonuniform Sampling}
Let us consider sampling based on a nonuniform partitioning that is represented using a binary tree. The approach examined here is inspired by the general framework given in \cite{chou1989optimal} for optimizing tree-structures for various tasks, and is also influenced by the discrete Lagrangian optimization approach \cite{everett1963generalized} and its application in coding \cite{shoham1988efficient,ortega1998rate}. 

\begin{figure}[]
	\centering
	{\subfloat[]{\label{fig:tree_structured_partitioning__full_tree}\includegraphics[width=0.24\textwidth]{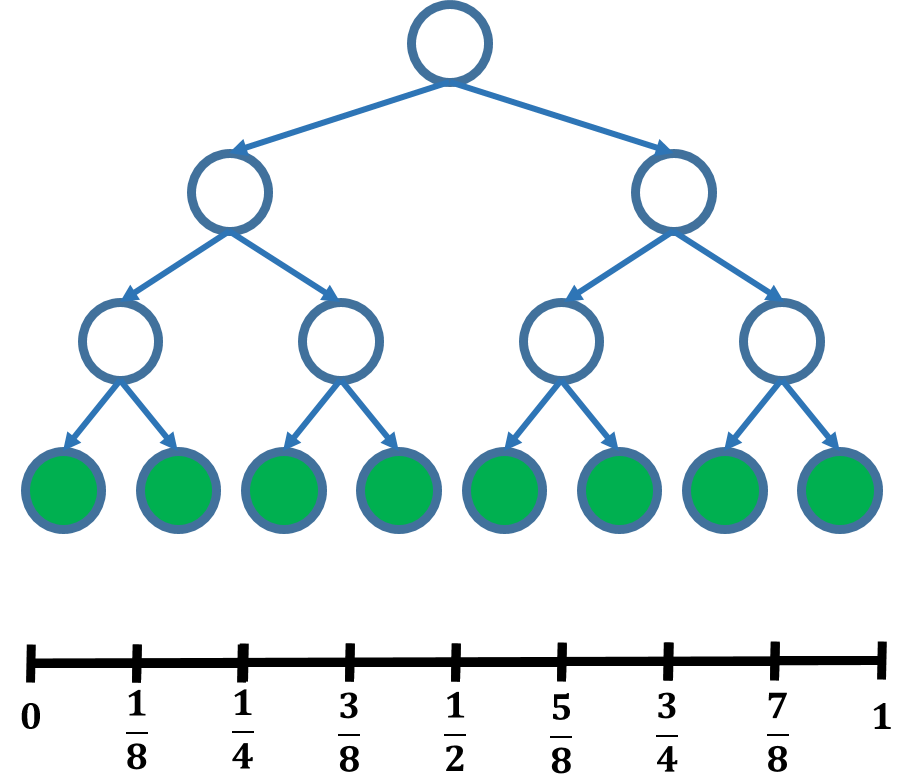}}}\\
	{\subfloat[]{\label{fig:tree_structured_partitioning__one_pruning}\includegraphics[width=0.24\textwidth]{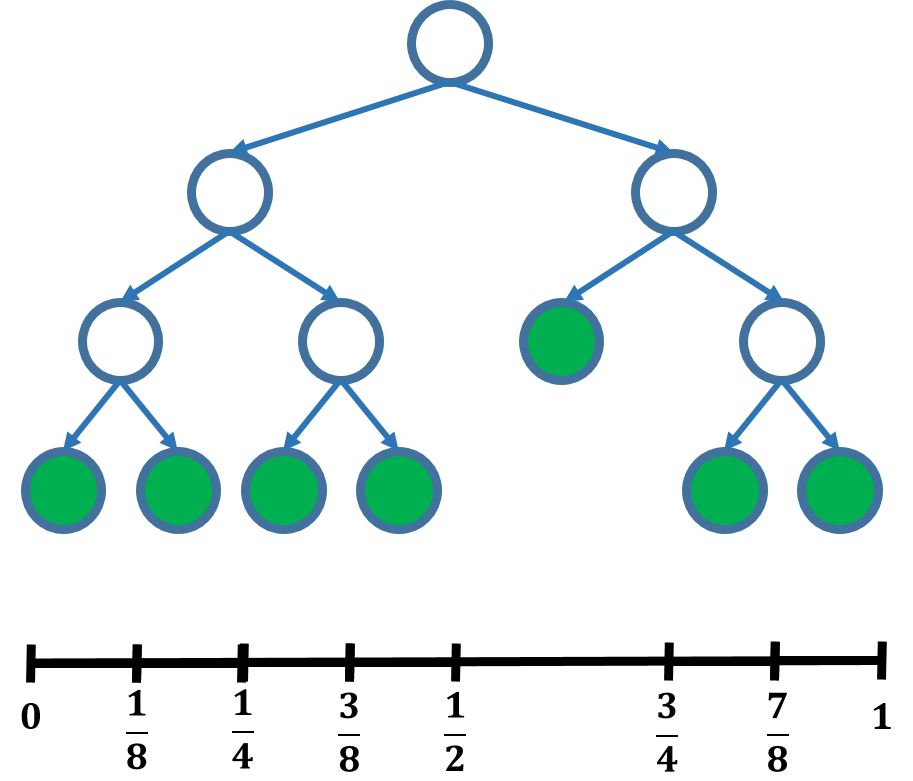}}}
	{\subfloat[]{\label{fig:tree_structured_partitioning__few_prunings}\includegraphics[width=0.24\textwidth]{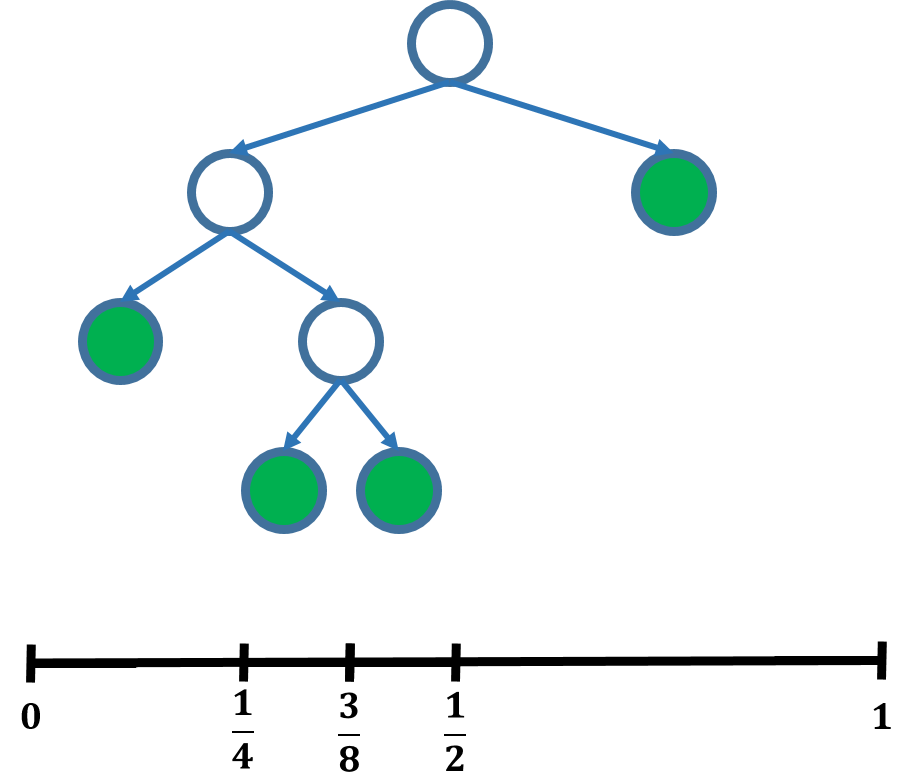}}}
	\caption{Segmentations of the $ [0,1) $ interval produced by binary trees. The leaves, which determine the partitioning, are colored in green. (a) A full binary tree of depth 3 and the corresponding uniform segmentation into 8 sub-intervals. (b) A tree obtained by a single pruning of the full tree, thus the partitioning includes 7 sub-intervals. (c) A tree obtained by several prunings resulting in 4 leaves/segments. } 
	\label{Fig:tree-structured sampling - partitioning demonstrations}
\end{figure}

The suggested approach relies on an initial tree, which is a full $ d $-depth binary-tree, representing a uniform segmentation of the interval $ \left[0,1\right) $ into $ 2^d $ sampling intervals of $ 2^{-d} $ length (see example in Fig. \ref{fig:tree_structured_partitioning__full_tree}). The segmentation of the interval $ \left[0,1\right) $ is described by the leaves of the binary tree: the interval location and length are defined by the leaf position in the tree, specifically, the tree-level where the leaf resides in determines the interval length.
The examined nonuniform segmentations are induced by all the trees obtained by repeatedly pruning neighboring-leaves having the same parent node (examples are given in Figures \ref{fig:tree_structured_partitioning__one_pruning}-\ref{fig:tree_structured_partitioning__few_prunings}). The initial $ d $-depth full-tree together with all its pruned subtrees form the set of relevant trees, denoted here as $ \mathcal{T}_d $.

The leaves of a tree $ T \in \mathcal{T}_d $ form a set denoted as $ L(T) $, where the number of leaves is referred to as $ |L(T)| $. Accordingly, the tree $ T $ represents a (possibly) nonuniform partitioning of the $ [0,1) $ interval into $ |L(T)| $ sampling intervals. A leaf $ l\nolinebreak\in\nolinebreak L(T) $ resides in the $ h(l) $ level of the tree and corresponds to the interval $ \left[ a^{left}_{(l)} , a^{right}_{(l)}  \right) $ of length $\Delta\left(l\right)\nolinebreak=\nolinebreak 2^{-h(l)}$. Following the analysis in section \ref{sec:Analysis for One-Dimensional Signals}, the optimal sample corresponding to the leaf $ l\in L(T) $ is expressed via (\ref{eq:optimal sample}) as
\begin{IEEEeqnarray}{rCl}
	\label{eq:tree-structured sampling - optimal sample}
	{\varphi _{(l)} ^{opt}} = \frac{1}{\Delta\left(l\right)} \mathop \int \limits_{a^{left}_{(l)}}^{a^{right}_{(l)}} \varphi \left( t \right)dt .
\end{IEEEeqnarray}
Consequently, as the tree leaves correspond to a segmentation of the $ [0,1) $ interval, the sampling MSE induced by the tree $ T \in \mathcal{T}_d $ is calculated based on the leaves, $L(T)$, via
\begin{IEEEeqnarray}{rCl}
	\label{eq:tree-structured sampling - sampling MSE for a tree}
	\errcost \left( T \right) = \mathop\sum_{l\in L(T)} {  \mathop \int \limits_{a^{left}_{(l)}}^{a^{right}_{(l)}}  {\left(  \varphi\left(t\right) - {\varphi _{(l)} ^{opt}}  \right) ^2 }  dt  } .
\end{IEEEeqnarray}

For a signal $ \varphi(t) $ and a budget of $ N $ samples, one can formulate the optimization of a tree-structured nonuniform sampling as
\begin{IEEEeqnarray}{rCl}
	\label{eq:tree-structured sampling - constrained optimization}
	\begin{aligned}
		& \underset{T\in\mathcal{T}_d}{\text{minimize}}
		& & \errcost \left( T \right) \\
		& \text{subject to}
		& & |L(T)| = N , 
	\end{aligned}
\end{IEEEeqnarray}
i.e., the optimization searches for the tree with $ N $ leaves that provides minimal sampling MSE. The unconstrained Lagrangian form of (\ref{eq:tree-structured sampling - constrained optimization}) is 
\begin{IEEEeqnarray}{rCl}
	\label{eq:tree-structured sampling - unconstrained Lagrangian optimization}
	\underset{T\in\mathcal{T}_d}{\min} \left\lbrace \errcost \left( T \right) + \mu |L(T)| \right\rbrace, 
\end{IEEEeqnarray}
where $ \mu \ge 0 $ is a Lagrange multiplier that reflects the constraint $|L(T)|=N$. However, it should be noted that due to the discrete nature of the problem such $ \mu $ does not necessarily exist for any $N$ value (for details see, e.g., \cite{chou1989optimal,everett1963generalized}).
The problem (\ref{eq:tree-structured sampling - unconstrained Lagrangian optimization}) can also be written as
\begin{IEEEeqnarray}{rCl}
	\label{eq:tree-structured sampling - unconstrained Lagrangian optimization - explicit}
	\underset{T\in\mathcal{T}_d}{\min} \left\lbrace \mathop\sum_{l\in L(T)} {  \mathop \int \limits_{a^{left}_{(l)}}^{a^{right}_{(l)}}  {\left(  \varphi\left(t\right) - {\varphi _{(l)} ^{opt}}  \right) ^2 }  dt  } + \mu |L(T)| \right\rbrace.
\end{IEEEeqnarray}
Note that, due the non-intersecting sampling intervals, the contribution of a leaf, $ l \in L(T) $, to the Lagrangian cost is 
\begin{IEEEeqnarray}{rCl}
	\label{eq:tree-structured sampling - leaf Lagrangian cost}
	C\left(l\right) = \mathop \int \limits_{a^{left}_{(l)}}^{a^{right}_{(l)}}  {\left(  \varphi\left(t\right) - {\varphi _{(l)} ^{opt}}  \right) ^2 }  dt  + \mu,
\end{IEEEeqnarray}	
evaluated for the corresponding sampling interval.

The discrete optimization problem (\ref{eq:tree-structured sampling - unconstrained Lagrangian optimization - explicit}) of finding the optimal tree for a given signal and a Lagrange multiplier $ \mu $ is addressed by the following procedure. 
Start from the full $d$-depth tree and determine the corresponding sampling intervals and their optimal samples, squared errors, and contributions to the Lagrangian cost (\ref{eq:tree-structured sampling - leaf Lagrangian cost}). Go through the tree levels from bottom and up, in each tree level find the pairs of neighboring leaves having the same parent node and evaluate the pruning condition: if 
\begin{IEEEeqnarray}{rCl}
	\label{eq:tree-structured sampling - leaf pruning condition}
	C\left(\text{left child}\right) + C\left(\text{right child}\right) > C\left(\text{parent}\right)
\end{IEEEeqnarray}	
is true, then prune the two leaves -- implying that two sampling intervals are merged to form a single interval of double length (thus, the total samples in the partitioning is reduced by one). If the condition (\ref{eq:tree-structured sampling - leaf pruning condition}) is false, then the two leaves (and the corresponding sampling intervals) are kept.
This procedure is continued until reaching a level where no pruning is done, or when getting to the tree root.

\section{Differential Coding of Monotonic Sequences of Integers}
\label{appendix:Differential Coding of Monotonic Sequences of Integers}

Following the practical compression method suggested in Subsection \ref{subsec:Compression based on the Proposed Sampling Method}, we elaborate here on the coding procedure of the segmentation and extrema time points. 
Let us consider, for example, the sequence of time points $ \left\lbrace a_i^{opt} \right\rbrace_{i=1}^{N-1} $, describing the nonuniform segmentation. In practice, we approximate the continuous time interval $ \left[ 0, 1 \right) $ using a discrete uniform grid in a high resolution corresponding to the small interval length of $ \Delta_{U} $. Accordingly, the to-be-encoded time-points practically obey $  a_i^{opt} = \tau_i \Delta_{U} $ where $ \tau_i \in \left\lbrace 0,1,..., \lfloor { \frac{1}{\Delta_{U}} } \rfloor \right\rbrace $ is the appropriate integer satisfying the relation. Since $ \Delta_{U} $ is known to the reconstruction process, we can accurately infer the sequence of time points $ \left\lbrace a_i^{opt} \right\rbrace_{i=1}^{N-1} $ from the integer sequence $ \left\lbrace \tau_i \right\rbrace_{i=1}^{N-1} $. Noting that $ \left\lbrace \tau_i \right\rbrace_{i=1}^{N-1} $ is monotonic in the sense that $ \tau_i \le \tau_{i+1} $ for any $ i=1,...,N-2 $.
Similarly, the extrema time points $ \left\lbrace x_j \right\rbrace_{j=1}^{J} $ can be also translated to a corresponding sequence of monotonic integers $ \left\lbrace \tilde{\tau}_j \right\rbrace_{j=1}^{J} $.

Let us overview the coding procedure for the sequence $ \left\lbrace \tau_i \right\rbrace_{i=1}^{N-1} $. We define the sequence of differences as $ \left\lbrace \tau_{diff,i} \right\rbrace_{i=1}^{N-1} $ where $\tau_{diff,1} = \tau_{1}$ and $ \tau_{diff,i} =  \tau_i - \tau_{i-1} $ for $ i = 2,...,N-1 $. 
Assume that we have an integer amount of $ b_{diff} $ bits for representing the basic encoding symbol. Since using $ b_{diff} $ bits one can describe a symbol with values in the integer range $ \left\lbrace 0,1,..., 2^{b_{diff}}-1 \right\rbrace $, we propose to encode each element of $ \left\lbrace \tau_{diff,i} \right\rbrace_{i=1}^{N-1} $ as follows. For $ i=1,...,N-1 $, if $ \tau_{diff,i} \le 2^{b_{diff}}-2 $ than it is encoded using a single $ b_{diff} $-bit symbol; otherwise, $\tau_{diff,i}$ can be encoded using $ 1 + \lfloor{ \frac{ \tau_{diff,i} }{2^{b_{diff}}-1} }\rfloor $ symbols of $ b_{diff} $ bits, where these symbols correspond to the value $ 2^{b_{diff}}-1 $, except to the last symbol that represents the value of $ \tau_{diff,i} - \left( 2^{b_{diff}}-1 \right) \cdot \lfloor{ \frac{ \tau_{diff,i} }{2^{b_{diff}}-1} }\rfloor $.

Evidently, the difference values $ \left\lbrace \tau_{diff,i} \right\rbrace_{i=1}^{N-1} $ are more likely to be small. Hence, while setting a too large $ b_{diff} $ will minimize the total number of coding symbols used, the total bit-cost will not be the minimal achievable by the above suggested procedure. Consequently, we precede the coding by finding the best $ b_{diff} $ in terms of minimal bit-cost for representing the given difference sequence. This search starts at the maximal value of  $ b_{diff} = \lceil { \log_2 \left( { \max_i \left\lbrace {\tau_{diff,i}} \right\rbrace } \right) } \rceil $ and continues to lower values, that for each an estimate of the total number of symbols and bits required is computed, without needing an explicit coding of the sequence. The best value found for $  b_{diff} $ is encoded using a small number of bits (in our experiments we used 4 bits) and associated with the encoded sequence of differences.

\section*{Acknowledgment}
The authors thank the reviewers for very pertinent and useful comments.

\ifCLASSOPTIONcaptionsoff
  \newpage
\fi



\bibliographystyle{IEEEtran}
\bibliography{IEEEabrv,high_rate_sampling__refs}
%

%
%

%





\end{document}